\documentclass[11pt,reqno,UTF8,twoside]{amsart}

\usepackage{mathrsfs}
\usepackage{amsfonts,amssymb,amsmath,amsthm,mathtools,bm}
\usepackage{cite}
\usepackage{geometry}
\usepackage{indentfirst}
\usepackage{setspace}
\usepackage{enumitem}
\usepackage{appendix}
\usepackage{graphicx,float,booktabs,array}
\usepackage{tikz}
\usetikzlibrary{calc,patterns,decorations.pathreplacing,arrows.meta,positioning}
\usepackage{caption}
\usepackage[
  colorlinks=true,
  linkcolor=magenta,
  citecolor=cyan,
  hypertexnames=false
]{hyperref}

\allowdisplaybreaks[2]
\makeatletter
\@namedef{subjclassname@2020}{\textup{2020} Mathematics Subject
  Classification}
\makeatother

\newtheorem{theorem}{Theorem}[section]
\newtheorem{lemma}[theorem]{Lemma}
\newtheorem{proposition}[theorem]{Proposition}
\newtheorem{corollary}[theorem]{\bfseries Corollary}

\theoremstyle{definition}
\newtheorem{definition}[theorem]{Definition}
\numberwithin{equation}{section}
\newcommand{\bbZ}{\mathbb Z}
\newcommand{\bbR}{\mathbb R}
\newcommand{\bbC}{\mathbb C}
\newcommand{\cH}{\mathcal H}
\newcommand{\cK}{\mathcal K}
\newcommand{\cF}{\mathcal F}
\newcommand{\one}{\mathbf 1}
\newcommand{\eps}{\varepsilon}
\newcommand{\dist}{\operatorname{dist}}
\newcommand{\supp}{\operatorname{supp}}
\newcommand{\Leb}{\operatorname{Leb}}
\newcommand{\diag}{\operatorname{diag}}
\newcommand{\Tr}{\operatorname{Tr}}
\newcommand{\Ran}{\operatorname{Ran}}
\newcommand{\Span}{\operatorname{span}}
\newcommand{\HS}{\mathcal J_2}
\newcommand{\TC}{\mathcal J_1}
\newcommand{\im}{\operatorname{Im}}
\newcommand{\la}{\langle}
\newcommand{\ra}{\rangle}

\title[]
{All-Disorder Localization for the One-Dimensional Anderson Model with Power-Law Hopping}

\author[]{Huajian Jiang}
\address[H.J.]{School of Mathematics, Sichuan University, Chengdu 610064, China}
\email{huajianjiang@scu.edu.cn}

\author[]{Yunfeng Shi}
\address[Y.S.]{School of Mathematics, Sichuan University, Chengdu 610064, China}
\email{yunfengshi@scu.edu.cn}

\date{November 2, 2026}

\begin{document}



\begin{abstract}
We consider the one-dimensional random operator
\[
(H_{\lambda,\omega}u)(n)
=
\sum_{m\ne n}\frac{u(m)}{|n-m|^r}
+\lambda\omega_nu(n),
\]
where the random variables $\omega_n$ are independent and uniformly distributed on $[-1,1]$
(i.e., \mbox{$\omega_n\stackrel{\mathrm{i.i.d.}}{\sim}\operatorname{Unif}[-1,1]$}).
For every $\lambda>0$ and $r>3/2$, we prove that $H_{\lambda,\omega}$ almost surely has pure point spectrum and admits an orthonormal eigenbasis $\{\psi_j\}$ satisfying
\[
\sum_{n\in\bbZ}(1+|n|)^{2\tau}|\psi_j(n)|^2<\infty
\qquad
\text{for every }j\text{ and }0\le\tau<r-\frac32.
\]
These conclusions establish spectral localization, thereby rigorously confirming the prediction of earlier analytical and numerical studies in the physics literature, which identified $r=3/2$ as the marginal exponent.
\end{abstract}

\maketitle

\section{Introduction}
\label{sec:model}

Consider the random operator
\begin{equation}
(H_{\lambda,\omega}u)(n)
=
\sum_{m\ne n}|n-m|^{-r}u(m)+\lambda\omega_nu(n),
\qquad
\omega_n\stackrel{\mathrm{i.i.d.}}{\sim}\operatorname{Unif}[-1,1].
\label{eq:model}
\end{equation}
Throughout the paper, we fix $r>3/2$ and $\lambda>0$. Write
\begin{equation}
T_r(n,m)=\one_{n\ne m}|n-m|^{-r},
\qquad
\alpha=r-\frac32>0.
\label{eq:alpha}
\end{equation}
The study of disorder-induced localization goes back to Anderson's seminal work \cite{Anderson1958}. In one dimension, long-range hopping can change the localization picture. Analytical and numerical studies of models with power-law hopping and i.i.d.\ uniformly distributed diagonal disorder predict a localization--delocalization transition for certain states as the disorder strength varies when $1<r<3/2$, and identify $r=3/2$ as the marginal exponent \cite{deMouraEtAl2005,MalyshevEtAl2004,RodriguezEtAl2003}. Recent work also reports semilocalized ground-state behavior in momentum space at weak disorder for $1<r<3/2$ \cite{BahovadinovEtAl2026}.

This value can be understood through a scaling argument \cite{RodriguezEtAl2003,deMouraEtAl2005}. For a chain of $N$ sites with periodic boundary conditions and $1<r<3$, the disorder-free level spacing at the upper band edge scales as $N^{1-r}$, whereas the typical magnitude of the matrix elements of the random potential between normalized plane waves scales as $\lambda N^{-1/2}$. The ratio of this magnitude to the level spacing is therefore of order $\lambda N^{r-3/2}$, identifying $r=3/2$ as the marginal exponent.

Mathematical results for model~\eqref{eq:model} establish localization for $r>1$ at sufficiently large disorder through the fractional-moment method of Aizenman and Molchanov \cite{AizenmanMolchanov1993}. Aizenman \cite{Aizenman1994Weak} proved localization at weak disorder in certain energy regimes, with an extension to polynomial hopping for $r>2$. Shi \cite{Shi2021} established power-law localization by multiscale analysis for H\"older continuous single-site distributions, assuming sufficiently large $r$ and sufficiently large disorder. See also \cite{AloisioEtAl2026,HislopMatosRojasMolina2026} for further results on random operators with power-law hopping at large disorder. To the best of our knowledge, localization throughout the spectrum of model~\eqref{eq:model} at arbitrary nonzero disorder strength was previously known only for $r>8$, through the work of Jak\v{s}i\'c and Molchanov \cite{JaksicMolchanov1999}.

For $r>3/2$, all states are predicted to be localized at arbitrary nonzero disorder strength \cite{deMouraEtAl2005}. We extend the result of Jak\v{s}i\'c and Molchanov to this physically predicted range: for every fixed $\lambda>0$ and $r>3/2$, we prove that model~\eqref{eq:model} almost surely exhibits spectral localization throughout the spectrum and admits an orthonormal eigenbasis satisfying weighted $\ell^2$ bounds. Our result rigorously confirms the above physical prediction in the spectral sense.

\subsection{Main result}

Since $r>1$ and
$\sum_{k\ne0}|k|^{-r}=2\sum_{k=1}^{\infty}k^{-r}<\infty$,
the operator $H_{\lambda,\omega}$ is bounded and self-adjoint, with
$\|H_{\lambda,\omega}\|\le 2\sum_{k=1}^{\infty}k^{-r}+\lambda$.
For $\tau\ge0$, set $\langle n\rangle=1+|n|$ and define
\[
\ell^2_\tau(\bbZ)
=
\left\{u:\sum_{n\in\bbZ}\langle n\rangle^{2\tau}|u(n)|^2<\infty\right\}.
\]
For a self-adjoint operator $H$, let $\mathcal H_{\mathrm c}(H)$ denote its
continuous spectral subspace.

\begin{theorem}\label{thm:main}
For the model \eqref{eq:model}, assume $r>3/2$. For each fixed $\lambda>0$,
\[
\mathcal H_{\mathrm c}(H_{\lambda,\omega})=\{0\}
\]
with probability one. Moreover, $H_{\lambda,\omega}$ admits an orthonormal
eigenbasis $\{\psi_j\}$ satisfying
\begin{equation}
\psi_j\in\bigcap_{0\le\tau<r-3/2}\ell^2_\tau(\bbZ)
\qquad\text{for every }j.
\label{eq:weighted-basis}
\end{equation}
In particular, for every $0\le\tau<r-\frac32$,
\[
|\psi_j(n)|\le C_{j,\tau,\omega}\langle n\rangle^{-\tau}.
\]
\end{theorem}

\subsection{Main ideas of the proof}
Our proof combines barrier decoupling with a block fractional-moment argument. Two difficulties arise in the present setting. Long-range hopping couples intervals across the barriers, and the disorder strength may be arbitrarily small. We therefore need quantitative estimates on the effective couplings and a contraction mechanism that works for each fixed $\lambda>0$.

For each fixed reference energy, we use the gap-opening theorem for periodic potentials of Jak\v{s}i\'c--Molchanov \cite[Theorem~4.1]{JaksicMolchanov1999} to construct barrier blocks with a spectral gap. Auxiliary random marks allow us to select these barriers while preserving conditional independence of the disorder on the intervals between the barriers. This independence is needed for the subsequent spectral averaging.

We then apply the Schur complement to eliminate the barrier variables. The Laplace representation of $T_r(n,m)$ yields factorization estimates for couplings across spatial cuts, with decay exponent $r-3/2$. Together with resolvent estimates away from the spectrum, these bounds control the effective block couplings. We factor these couplings through Hilbert--Schmidt operators independent of the spectral parameter. Our multi-site spectral-averaging estimate uses the Hilbert--Schmidt weak-$L^1$ resolvent bound of \cite[Lemma~3.1]{AizenmanEtAl2006}, which also underlies their fractional-moment analysis of continuum operators. Here the factors are chosen so that the averaging coefficients tend to zero as the intervals between the barriers recede to infinity. The resulting block fractional-moment recursion is contractive outside a finite set of blocks for every fixed $\lambda>0$.

The same weak-$L^1$ estimate also provides the resolvent boundary limits needed to obtain weighted Green-function bounds. A Simon--Wolff argument \cite{SimonWolff1986} then yields pure point spectrum and an orthonormal eigenbasis with the stated weighted decay.

Our argument builds on the barrier method of Simon--Spencer \cite{SimonSpencer1989} and its development for long-range operators by Jak\v{s}i\'c--Molchanov \cite{JaksicMolchanov1999}. Simon and Spencer used trace-class perturbations to exclude absolutely continuous spectrum, while Jak\v{s}i\'c and Molchanov obtained pure point spectrum through a cluster expansion of the resolvent. Here we follow \cite{JaksicMolchanov1999} in constructing barriers from periodic spectral gaps, but modify the selection procedure to preserve conditional independence. We then use a Schur-complement reduction and a block fractional-moment argument in place of their resolvent cluster expansion.

\subsection{Organization of the paper}

The rest of the paper is organized as follows.
Section~\ref{sec:preliminaries} fixes the notation and collects the
spectral-averaging, factorization, resolvent, and effective-coupling
estimates. It includes the spatial-cut calculation that yields the
exponent $r-3/2$; the technical proofs are given in
Appendices~\ref{app:averaging-proofs}--\ref{app:resolvent-proofs}.
Section~\ref{sec:periodic} establishes finite-volume spectral gaps for
suitable periodic potentials and uses auxiliary random marks to select
barriers while preserving conditional independence of the disorder on
the intervals between them.
Section~\ref{sec:schur-factorization} carries out the Schur-complement
reduction and derives recursive fractional-moment bounds.
Section~\ref{sec:terminal} establishes weighted Green-function bounds
and proves spectral localization by rank-one spectral averaging.
Appendix~\ref{app:measurability} proves the measurable selection result
used to choose the factorization operators in Section~\ref{sec:schur-factorization}.

\section{Notation and key estimates}
\label{sec:preliminaries}

We collect the notation and estimates used in the barrier construction
and the fractional-moment argument. The spatial-cut computation is included
here; the technical proofs are given in
Appendices~\ref{app:averaging-proofs}--\ref{app:resolvent-proofs}.

\subsection{Notation and conventions}
\label{sec:notation}

\begin{itemize}
\item We set $\mathbb N_0=\{0,1,2,\ldots\}$ and write $\bigsqcup$ for a disjoint union. For finite $A\subset\bbZ$, $|A|$ denotes its cardinality. For subsets $A,B\subset\bbZ$, we set
\[
\dist(A,B):=\inf\{|x-y|:x\in A,\ y\in B\}.
\]

\item We write $\lfloor t\rfloor$ and $\lceil t\rceil$ for the greatest integer not exceeding $t$ and the least integer not less than $t$, respectively.

\item The symbol $\one_A$ denotes the indicator of $A$. For a set $S\subset\bbR$, $S^\circ$ denotes its interior.

\item The vector $\delta_x$ is the canonical basis vector at $x$. For $A\subset\bbZ$, let $P_A$ be the orthogonal projection onto $\ell^2(A)$, viewed as a subspace of $\ell^2(\bbZ)$. For an operator $H$ on $\ell^2(\bbZ)$, write $H_A=P_AHP_A$ for its compression to $\ell^2(A)$. For $X,Y\subset\bbZ$, set $T_{XY}:=P_XT_rP_Y$.

\item We write $A(n,m):=\langle\delta_n,A\delta_m\rangle$. For $a=(a_x)_{x\in X}$, let $\diag a$ denote the diagonal operator on $\ell^2(X)$ defined by $(\diag a)\delta_x=a_x\delta_x$.

\item With $\langle n\rangle=1+|n|$, for $\tau\ge0$ and $A\subset\bbZ$ we set
\[
\begin{aligned}
\ell^2_\tau(A)
&=\left\{u:A\to\bbC:\sum_{n\in A}\langle n\rangle^{2\tau}|u(n)|^2<\infty\right\},\\
\|u\|_{\ell^2_\tau(A)}^2
&=\sum_{n\in A}\langle n\rangle^{2\tau}|u(n)|^2.
\end{aligned}
\]

\item For a self-adjoint operator $H$, $\sigma(H)$, $\sigma_{\mathrm p}(H)$, $\sigma_{\mathrm{ess}}(H)$, and $\rho(H)$ denote its spectrum, point spectrum, essential spectrum, and resolvent set. We denote by $\sigma_{\mathrm{disc}}(H)$ the discrete spectrum of $H$, consisting of isolated eigenvalues of finite multiplicity. For a Borel set $S\subset\bbR$, $P_S(H)=\one_S(H)$ denotes its spectral projection onto $S$. We denote by $P_{\mathrm c}(H)$ the orthogonal projection onto the continuous spectral subspace $\mathcal H_{\mathrm c}(H)$.

\item Inner products are linear in the second argument. For vectors $u,v$, $|u\rangle\langle v|$ denotes the operator $w\mapsto u\langle v,w\rangle$.

\item We write $\mathcal B(\cH_1,\cH_2)$ for the bounded operators from $\cH_1$ to $\cH_2$, and $\mathcal B(\cH)=\mathcal B(\cH,\cH)$. The symbols $D(A)$, $A^*$, and $\Ran A$ denote the domain, adjoint, and range of $A$, respectively. For an orthogonal projection $P$, $P^\perp=I-P$.

\item Unless otherwise specified, $\lVert\cdot\rVert$ denotes the Hilbert-space norm for vectors and the operator norm for operators; $\|v\|_\infty=\sup_x|v(x)|$.

\item For a bounded operator $Q$, set $\im Q=(Q-Q^*)/(2i)$. We write $\im Q<0$ if $\langle u,(\im Q)u\rangle<0$ for every nonzero $u$.

\item We write $\Leb$ for Lebesgue measure, $\mathbb P$ for probability, and $\mathbb E$ for expectation, with the conditional-law conventions in Sections~\ref{sec:schur-factorization} and~\ref{sec:terminal} specified there. For a random variable $X$ and a sub-$\sigma$-algebra $\mathcal G$, $\mathcal L(X\mid\mathcal G)$ denotes a conditional law and $\mathbb E(X\mid\mathcal G)$ denotes conditional expectation. We abbreviate $\mathcal L(X\mid Y):=\mathcal L(X\mid\sigma(Y))$, where $\sigma(Y)$ is the $\sigma$-algebra generated by $Y$.

\item We use the unitary Fourier transform
\[
(\mathcal Fu)(\varphi)=\frac1{\sqrt{2\pi}}\sum_{n\in\bbZ}u(n)e^{-in\varphi},
\qquad -\pi\le\varphi<\pi,
\]
from $\ell^2(\bbZ)$ onto $L^2([-\pi,\pi])$. For a bounded measurable function $f$, $M_f$ denotes multiplication by $f$.

\item We write $K\Subset U$ if $K$ is compact and $K\subset U$.

\item We write $X\asymp Y$ when $cY\le X\le CY$, with constants independent of the relevant scale parameters. Constants denoted by $C,C_1,C_2,\ldots$ may change from line to line; subscripts indicate their allowed dependence.
\end{itemize}

\subsection{Schatten classes}
For Hilbert spaces $\cK_1,\cK_2$ and a compact operator
$A:\cK_2\to\cK_1$, let $s_j(A)$ be the singular values of
$A$, i.e. the nonzero eigenvalues of $|A|=(A^*A)^{1/2}$, counted with multiplicity, arranged in nonincreasing order, and padded with zeros when necessary. For $p=1,2$, the Schatten class $\mathcal J_p(\cK_2,\cK_1)$ consists of the compact operators $A:\cK_2\to\cK_1$ for which
\[
\|A\|_{\mathcal J_p}:=\left(\sum_j s_j(A)^p\right)^{1/p}<\infty.
\]
We write $\mathcal J_p$ when the spaces are clear, and set $\mathcal J_p(\cK)=\mathcal J_p(\cK,\cK)$. The classes $\TC$ and $\HS$ are the trace class and Hilbert--Schmidt
class, respectively. These are Banach spaces, and
$\|A^*\|_{\mathcal J_p}=\|A\|_{\mathcal J_p}$ and $\|A\|\le\|A\|_{\mathcal J_p}$.
If $L,R$ are bounded and the product is defined, then
$\|LAR\|_{\mathcal J_p}\le\|L\|\|A\|_{\mathcal J_p}\|R\|$.
For the basic properties of Schatten classes see
\cite[Chapters~1--2]{Simon2005TraceIdeals}.

For any orthonormal basis $(e_j)$ of $\cK_2$,
\[
\|A\|_{\HS}^2=\sum_j\|Ae_j\|^2.
\]
Moreover, $\|A\|_{\TC}=\operatorname{Tr}|A|$.
For a positive operator $A\in\TC(\cK)$, $\|A\|_{\TC}=\operatorname{Tr}A$.
For vectors $u,v$, we have
$\||u\rangle\langle v|\|_{\TC}=\|u\|\|v\|$.
We use the Schatten--H\"older inequality
\[
\|F^*G\|_{\TC}\le\|F\|_{\HS}\|G\|_{\HS}
\]
for Hilbert--Schmidt operators $F:\cK_1\to\cH$ and $G:\cK_2\to\cH$; see
\cite[Theorem~2.8]{Simon2005TraceIdeals}.

\subsection{Spectral averaging}
\label{sec:averaging}

We recall the Hilbert--Schmidt weak-$L^1$ estimate of
\cite[Lemma~3.1]{AizenmanEtAl2006} in the form needed below, and state
the resulting multi-site fractional-moment bound. The proofs of
Lemmas~\ref{lem:weakL1}, \ref{lem:density}, and \ref{lem:multi-site}
are given in Appendix~\ref{app:averaging-proofs}.

Our convention is that a densely defined operator $A$ on a Hilbert space $\cH$ is dissipative if $\im\langle u,Au\rangle\ge0$ for every $u\in D(A)$. For operators $A,B$ on $\cH$, we write $A\subseteq B$ if $D(A)\subseteq D(B)$ and $Bu=Au$ for every $u\in D(A)$. A dissipative operator is maximally dissipative if it has no proper dissipative extension on $\cH$; see \cite[Remark following Lemma~3.1]{AizenmanEtAl2006}.

\begin{lemma}
\label{lem:weakL1}
Let $A$ be maximally dissipative on a separable Hilbert space $\cH$, and let
$M_2:\cK_2\to\cH$ and $M_1:\cH\to\cK_1$ be Hilbert--Schmidt operators.
The operators $M_1(A-t+i\eps)^{-1}M_2$ converge in
$\HS(\cK_2,\cK_1)$ as $\eps\downarrow0$ for Lebesgue-almost every
$t\in\bbR$. Denote this limit by
\begin{subequations}\label{eq:weakL1}
\begin{equation}
M_1(A-t+i0)^{-1}M_2
:=
\lim_{\eps\downarrow0}M_1(A-t+i\eps)^{-1}M_2.
\label{eq:weakL1-limit}
\end{equation}
Then, for $u>0$,
\begin{equation}
\Leb\left\{t:\|M_1(A-t+i0)^{-1}M_2\|_{\HS}>u\right\}
\le
\frac{C_{\mathrm W}}{u}\|M_1\|_{\HS}\|M_2\|_{\HS}.
\label{eq:weakL1-tail}
\end{equation}
\end{subequations}
Here $C_{\mathrm W}>0$ is an absolute constant. If $A$ is self-adjoint, the same conclusions hold with
$+i\eps$ replaced by $-i\eps$.
\end{lemma}

\begin{definition}\label{def:phi}
Let $X$ be a finite or countable set, let $\cK$ be a Hilbert space, and let
$F:\ell^2(X)\to\cK$ be a bounded linear operator. Define the column functional
\begin{equation}
\Phi(F)
=
\left(\sum_{x\in X}\|F\delta_x\|_{\cK}^{4/3}\right)^{3/2}
\in[0,\infty].
\label{eq:Phi}
\end{equation}
Whenever $\Phi(F)<\infty$,
\begin{equation}
\|F\|_{\HS}^2
=\sum_{x\in X}\|F\delta_x\|_{\cK}^2
\le\Phi(F).
\label{eq:phi-hs}
\end{equation}
Indeed, set $c=(c_x)_{x\in X}$, where $c_x:=\|F\delta_x\|_{\cK}$.
Since $\{\delta_x\}_{x\in X}$ is the canonical orthonormal basis of $\ell^2(X)$,
$\|F\|_{\HS}^2=\|c\|_{\ell^2}^2$. If $\Phi(F)<\infty$, then
$c\in\ell^{4/3}(X)$ and $\|c\|_{\ell^2}\le\|c\|_{\ell^{4/3}}$,
which proves \eqref{eq:phi-hs}. In particular, $\Phi(F)<\infty$ implies
$F\in\HS(\ell^2(X),\cK)$.
\end{definition}

We use $H^1(\bbR)$ to denote the first-order Sobolev space
\[
H^1(\bbR)=\{f\in L^2(\bbR):f'\in L^2(\bbR)\},
\]
where $f'$ denotes the distributional, equivalently weak, derivative.

\begin{definition}\label{def:mixed-info}
Let $p$ be a probability density on $\bbR$. If $\sqrt p\in H^1(\bbR)$,
define its Fisher information by
\[
I(p):=4\int_{\bbR}|(\sqrt p)'(t)|^2\,dt
=\int_{\{p>0\}}\frac{|p'(t)|^2}{p(t)}\,dt.
\]
If $\sqrt p\notin H^1(\bbR)$, set $I(p)=+\infty$.
We say that $p$ has finite mixed Fisher information if there are
a finite or countable family of probability densities $p_\nu$ with
$I(p_\nu)<\infty$ and weights $\pi_\nu\ge0$ such that
\begin{equation}
p=\sum_\nu\pi_\nu p_\nu,
\qquad
\sum_\nu\pi_\nu=1,
\qquad
\sum_\nu\pi_\nu I(p_\nu)^{1/3}<\infty.
\label{eq:mixed-info}
\end{equation}
For the decomposition in \eqref{eq:mixed-info}, write
$M:=\sum_\nu\pi_\nu I(p_\nu)^{1/3}$ for its finite
mixed-information bound.
\end{definition}

\begin{lemma}\label{lem:density}
Let $a$ be a smooth real-valued function on $[-1,1]$ satisfying
$0<c\le a(t)\le C$ for all $t\in[-1,1]$.
Then the probability density
\[
q_a:=\frac{a\one_{(-1,1)}}{\int_{-1}^1a}
\]
satisfies \eqref{eq:mixed-info}. In particular, the uniform density
$\frac12\one_{(-1,1)}$ satisfies \eqref{eq:mixed-info}.
If $b$ is a nonzero real-valued smooth compactly supported function, then
$b^2/\int b^2$ has finite Fisher information.
\end{lemma}

\begin{lemma}
\label{lem:multi-site}
Let $X$ be finite, let $F:\ell^2(X)\to K$ be an operator into a Hilbert space $K$, let $Q$ be a deterministic matrix on $\ell^2(X)$ with $\im Q<0$, and let
$V_x$ be independent random variables. For each $x$, fix an admissible
decomposition of its density $p_x$ as in \eqref{eq:mixed-info}, and let
$M_x$ be the resulting mixed-information bound. Assume
$M_*:=\sup_{x\in X}M_x<\infty$. Then, for $0<s<2/3$,
\begin{equation}
\mathbb E\left\|F(Q+\lambda\diag V)^{-1}F^*\right\|_{\HS}^{s}
\le
C_s\lambda^{-s}M_*^{3s/2}\Phi(F)^s.
\label{eq:multi-site}
\end{equation}
Here $C_s$ depends only on $s$.
\end{lemma}

\subsection{Factorization across spatial cuts}
\label{sec:factorization}

The functional below measures the size of a coupling through the column
functionals of its factors. The half-line computation will give the
decay exponent $\alpha=r-3/2$ without a factor depending on the lengths
of the coupled intervals.
The proofs of Lemmas~\ref{lem:gamma-algebra} and
\ref{lem:crosscut-product} are given in
Appendix~\ref{app:factorization-proofs}.

\begin{definition}\label{def:gamma}
Let $B:\ell^2(I)\to\ell^2(J)$, where $I,J$ are finite or countable
coordinate sets. Define
\begin{equation}
\gamma(B)
=
\inf_{B=F^*G}\sqrt{\Phi(F)\Phi(G)},
\label{eq:gamma}
\end{equation}
where the infimum is taken over Hilbert spaces $\mathcal K$ and factorizations
\[
G:\ell^2(I)\to\mathcal K,
\qquad
F:\ell^2(J)\to\mathcal K,
\qquad
B=F^*G.
\]
We use the convention that the infimum over an empty set is $+\infty$.
\end{definition}

Set $p_0=2/3$. By Definition~\ref{def:phi},
\[
\Phi(F)^{p_0}=\sum_x\|F\delta_x\|^{4/3}.
\]
For a matrix $A$ on $\ell^2(\bbZ)$, set
\[
\|A\|_{\mathcal C}
:=\sum_{k\in\bbZ}\sup_{n\in\bbZ}|A(n,n-k)|.
\]

\begin{lemma}\label{lem:gamma-algebra}
Let $I,J$ be finite or countable coordinate sets and let
$B,B_1,B_2:\ell^2(I)\to\ell^2(J)$. Then
\begin{align}
\|B\|_{\TC}&\le\gamma(B),\label{eq:gamma-trace}\\
\gamma(B_1+B_2)^{p_0}&\le\gamma(B_1)^{p_0}+\gamma(B_2)^{p_0}.
\label{eq:gamma-quasi}
\end{align}
For a countable family $(B_j)$ with
$\sum_j\gamma(B_j)^{p_0}<\infty$, the sum $B=\sum_jB_j$ is well defined
and satisfies
\begin{equation}
\gamma(B)^{p_0}\le\sum_j\gamma(B_j)^{p_0}.
\label{eq:gamma-countable}
\end{equation}
If $A,B,C$ are operators on $\ell^2(\bbZ)$, then
\begin{equation}
\gamma(ABC)\le\|A\|_{\mathcal C}\,\gamma(B)\,\|C\|_{\mathcal C},
\label{eq:gamma-product}
\end{equation}
For coordinate subsets $I_0\subseteq I$ and $J_0\subseteq J$,
\begin{equation}
\gamma(P_{J_0}BP_{I_0})\le\gamma(B).
\label{eq:gamma-compression}
\end{equation}
If $I,J$ are finite, then
\begin{equation}
\gamma(B)\le(|I||J|)^{1/4}\|B\|_{\TC}.
\label{eq:gamma-finite}
\end{equation}
\end{lemma}

We now estimate $\gamma$ for the coupling induced by $T_r$ between two separated half-lines. For two half-lines
whose nearest sites are separated by $L\ge1$, write the left and right
coordinates as $-a$ and $L+b$, respectively, with $a,b\ge0$. Their distance
is $L+a+b$, so the cross-coupling matrix is
\[
K_L(a,b)=(L+a+b)^{-r},
\qquad a,b\ge0.
\]
We use the Laplace identity
\begin{equation}
x^{-r}=\frac1{\Gamma(r)}\int_0^\infty t^{r-1}e^{-xt}\,dt,
\qquad x>0,
\label{eq:laplace-power}
\end{equation}
where $\Gamma$ is the Euler gamma function. This gives the following factorization. Define
$F_L:\ell^2(\mathbb N_0)\to L^2(0,\infty)$ by
\[
(F_L\delta_n)(t)
=
\Gamma(r)^{-1/2}t^{(r-1)/2}e^{-(n+L/2)t},
\qquad t>0.
\]
For $a,b\ge0$, \eqref{eq:laplace-power} gives
\[
\langle F_L\delta_a,F_L\delta_b\rangle
=\frac1{\Gamma(r)}\int_0^\infty
 t^{r-1}e^{-(L+a+b)t}\,dt
=(L+a+b)^{-r}.
\]
Thus
\begin{equation}
K_L=F_L^*F_L,
\qquad
\|F_L\delta_n\|^2=(L+2n)^{-r}.
\label{eq:halfline-factor}
\end{equation}
Since $r>3/2$ and $L\ge1$,
\[
\sum_{n\ge0}(L+2n)^{-2r/3}
\le L^{-2r/3}+\int_0^\infty(L+2x)^{-2r/3}\,dx
\le C_rL^{1-2r/3}.
\]
Hence, by Definition~\ref{def:gamma},
\begin{equation}
\gamma(K_L)
\le\Phi(F_L)
=\left(\sum_{n\ge0}(L+2n)^{-2r/3}\right)^{3/2}
\le C_rL^{3/2-r}
=C_rL^{-\alpha}.
\label{eq:halfline-gamma}
\end{equation}
This is the first place where $r>3/2$ is used.

Since $K_L=F_L^*F_L$, we also have
\[
\|K_L\|_{\TC}=\Tr K_L
=\sum_{n\ge0}K_L(n,n)
=\sum_{n\ge0}(L+2n)^{-r}
\le C_rL^{1-r}.
\]
Let $P_x^-=\one_{(-\infty,x]}$ and $P_y^+=\one_{(y,\infty)}$. Define
\begin{equation}
g_\alpha(A)
=
\sup_{x\le y}(1+y-x)^\alpha
\max\left\{\gamma(P_x^-AP_y^+),\gamma(P_y^+AP_x^-)\right\}.
\label{eq:crosscut}
\end{equation}

For $x\le y$, write
\[
n=x-a,\qquad m=y+1+b,\qquad a,b\ge0.
\]
Then
\[
(P_x^-T_rP_y^+)(x-a,y+1+b)
=(1+y-x+a+b)^{-r}=K_{1+y-x}(a,b).
\]
By \eqref{eq:halfline-gamma},
\[
\gamma(P_y^+T_rP_x^-)
=\gamma(P_x^-T_rP_y^+)
=\gamma(K_{1+y-x})
\le C_r(1+y-x)^{-\alpha}.
\]
It follows that $g_\alpha(T_r)\le C_r$. If $D$ is diagonal, then
\[
P_x^-DP_y^+=P_y^+DP_x^-=0,
\qquad x\le y,
\]
so $g_\alpha(D)=0$.

\begin{lemma}\label{lem:crosscut-product}
Let $A,B$ be operators on $\ell^2(\mathbb Z)$ satisfying
\[
\|A\|_{\mathcal C},\ \|B\|_{\mathcal C},\ g_\alpha(A),\ g_\alpha(B)<\infty.
\]
Then
\begin{equation}
g_\alpha(AB)^{p_0}
\le
C_\alpha\left(
\|A\|_{\mathcal C}^{p_0}g_\alpha(B)^{p_0}
+g_\alpha(A)^{p_0}\|B\|_{\mathcal C}^{p_0}
\right).
\label{eq:crosscut-product}
\end{equation}
Here $C_\alpha=2^{\alpha p_0}$.
\end{lemma}

\subsection{Resolvent and effective coupling estimates}
\label{sec:inverse}
\label{sec:self-energy}

We next state the resolvent estimates and the bounds on effective
couplings used in the Schur-complement argument. The proofs of
Lemma~\ref{lem:pointwise-algebra}, Proposition~\ref{prop:uniform-inverse},
Corollary~\ref{cor:weighted-J}, and Proposition~\ref{prop:rectangular-self-energy}
are given in
Appendix~\ref{app:resolvent-proofs}.

Define
\[
\|A\|_{J_r}=\sup_{x,y}\langle x-y\rangle^r|A(x,y)|,
\qquad
S(A)=\max\left\{
\sup_x\sum_y|A(x,y)|,
\sup_y\sum_x|A(x,y)|
\right\}.
\]
Schur's test gives $\|A\|\le S(A)$.
Moreover, whenever $\|A\|_{\mathcal C}<\infty$,
\[
\sup_x\sum_y|A(x,y)|
\le\sum_k\sup_n|A(n,n-k)|=\|A\|_{\mathcal C},
\]
and the analogous column bound gives $S(A)\le\|A\|_{\mathcal C}$.

\begin{lemma}\label{lem:pointwise-algebra}
Let $r>1$, and let $A,B$ be bounded operators on $\ell^2(\mathbb Z)$ with
\mbox{$\|A\|_{J_r},\ \|B\|_{J_r}<\infty$}.
There exists $C_r>0$, depending only on $r$, such that
\begin{align}
S(A)&\le\|A\|_{\mathcal C}\le C_r\|A\|_{J_r},\label{eq:C-J}\\
S(A)&\le
C_r\|A\|^{\frac{2r-2}{2r-1}}
\|A\|_{J_r}^{\frac1{2r-1}},\label{eq:S-interp}\\
\|AB\|_{J_r}&\le
C_r\bigl(\|A\|_{J_r}S(B)+S(A)\|B\|_{J_r}\bigr).
\label{eq:J-product}
\end{align}
\end{lemma}

\begin{proposition}
\label{prop:uniform-inverse}
Let $\varnothing\ne\Lambda\subset\bbZ$, let
$v:\Lambda\to\bbR$ satisfy $\|v\|_\infty\le V_0$, and set
$H_\Lambda=T_{r,\Lambda}+\diag(v|_\Lambda)$. If $|z|\le E_0$ and
$\dist(z,\sigma(H_\Lambda))\ge\delta>0$, then there exists
$C>0$, depending only on $r,V_0,E_0,\delta$, such that
\begin{equation}
\|(H_\Lambda-z)^{-1}\|_{J_r}
+\|(H_\Lambda-z)^{-1}\|_{\mathcal C}
+g_\alpha((H_\Lambda-z)^{-1})
\le C.
\label{eq:uniform-inverse}
\end{equation}
\end{proposition}

\begin{corollary}\label{cor:weighted-J}
If $\|A\|_{J_r}<\infty$, then, for every $0\le\tau<r-1$,
\[
\|A\|_{\mathcal B(\ell^2_\tau)}\le C_{r,\tau}\|A\|_{J_r}.
\]
\end{corollary}

\begin{proposition}
\label{prop:rectangular-self-energy}
Assume $r>3/2$. Let $I,J\subset\bbZ$ be disjoint, nonempty finite
intervals, set $N_I=|I|$, $N_J=|J|$, and $d=\dist(I,J)\ge1$, and let
$B\subset\bbZ\setminus(I\cup J)$. If
$R\in\mathcal B(\ell^2(B))$ satisfies $\|R\|_{J_r}\le C_R$, then
\begin{equation}
\|T_{IB}RT_{BJ}\|_{\TC}
\le C_rC_R\sqrt{N_IN_J}\,d^{-r}.
\label{eq:rectangular-self-energy}
\end{equation}
Consequently, suppose that $R=(H_B-z)^{-1}$ satisfies the
hypotheses of Proposition~\ref{prop:uniform-inverse}, and define
\[
V_{IJ}(z):=T_{IJ}-T_{IB}(H_B-z)^{-1}T_{BJ}.
\]
Then
\begin{equation}
\|V_{IJ}(z)\|_{\TC}\le C\sqrt{N_IN_J}\,d^{-r},
\label{eq:V-two-bounds}
\end{equation}
\begin{equation}
\gamma(V_{IJ}(z))\le Cd^{-\alpha}.
\label{eq:V-gamma-bound}
\end{equation}
\end{proposition}

Proposition~\ref{prop:rectangular-self-energy} provides two bounds with
different roles. The trace-class estimate depends on the interval lengths
through the factor $(N_IN_J)^{1/2}$, while the $\gamma$ estimate controls couplings
across spatial cuts without a volume factor. They will be used to
factor the effective couplings in
Section~\ref{sec:schur-factorization}.

\section{Random barriers from periodic potentials}
\label{sec:periodic}
\label{sec:barriers}

We first apply the periodic gap-opening theorem of Jak\v{s}i\'c--Molchanov
\cite[Theorem~4.1]{JaksicMolchanov1999} to $T_r$. Define
\[
f_r(\varphi):=\sum_{n\ne0}|n|^{-r}e^{-in\varphi}
=2\sum_{n\ge1}n^{-r}\cos(n\varphi),
\qquad -\pi\le\varphi<\pi,
\]
so that $\mathcal FT_r\mathcal F^{-1}=M_{f_r}$.

\begin{theorem}
\label{thm:periodic-gap-input}
Let $K\Subset\sigma(T_r)^\circ$ be a compact interval. Then there exist
$\eps_0>0$ and $P_0$ such that, whenever
\[
0<\eps<\eps_0,\qquad
P\in\mathbb N,\quad P>P_0,\qquad
e_{k,P}=f_r(k\pi/P)\in K,
\]
there exists $\delta_{\eps,k,P}>0$ such that
\[
\sigma\!\left(T_r+\eps\,\one_{\{n\equiv0\ (\mathrm{mod}\ P)\}}\right)
\cap(e_{k,P},e_{k,P}+\delta_{\eps,k,P})=\varnothing.
\]
\end{theorem}

\begin{proof}
Clearly, $f_r$ is continuous, real, and even.
By \cite[\S7.12, p.~236]{Lewin1981}, $f_r$ is real analytic on $(0,\pi)$, and
\begin{equation}
f_r'(\varphi)
=
-\frac{\sin\varphi}{\Gamma(r)}
\int_0^\infty
\frac{t^{r-1}\sinh t}
     {(\cosh t-\cos\varphi)^2}\,dt
<0,
\qquad 0<\varphi<\pi.
\label{eq:symbol-monotone}
\end{equation}
Thus $f_r$ satisfies the regularity and monotonicity assumptions of
\cite[Theorem~4.1]{JaksicMolchanov1999}, and the stated compact-interval formulation follows.
\end{proof}

\begin{lemma}\label{lem:edge-spectrum}
Let $A=T_r+w$ with $r>1$, where $w:\bbZ\to\bbR$ is $P$-periodic. For
$j=0,\ldots,P-1$, set
\[
A_{-,j}:=A_{(-\infty,j]\cap\bbZ},
\qquad
A_{+,j}:=A_{[j,\infty)\cap\bbZ},
\]
and define
\[
\Sigma_{\rm edge}(A)
:=
\bigcup_{j=0}^{P-1}
\bigl(\sigma(A_{-,j})\cup\sigma(A_{+,j})\bigr).
\]
Suppose that $J\cap\sigma(A)=\varnothing$ and set
$I_L=[-2PL,2PL]\cap\bbZ$. If $K\Subset J$ and
\[
E_L\in\sigma(A_{I_L})\cap K,
\qquad
E_L\to E,
\]
then
\[
E\in\Sigma_{\rm edge}(A).
\]
Moreover,
\[
\#\bigl(\Sigma_{\rm edge}(A)\cap K\bigr)<\infty,
\]
and, for every $j=0,\ldots,P-1$,
\[
\sigma(A_{\pm,j})\cap K\subset\sigma_{\rm p}(A_{\pm,j}),
\qquad
\dim\ker(A_{\pm,j}-E)<\infty
\quad\bigl(E\in\sigma(A_{\pm,j})\cap K\bigr).
\]
\end{lemma}

\begin{proof}
Fix $j\in\bbZ$ and define
\[
B_j:=P_{(-\infty,j]\cap\bbZ}T_rP_{[j+1,\infty)\cap\bbZ},
\]
Then
\[
B_j(j-a,j+1+b)=(1+a+b)^{-r},\qquad a,b\ge0.
\]
Hence
\[
\|B_j\|_{\HS}^2
=\sum_{a,b\ge0}(1+a+b)^{-2r}
=\sum_{q\ge0}(q+1)^{1-2r}<\infty.
\]
With respect to
\[
\ell^2(\bbZ)=\ell^2((-\infty,j]\cap\bbZ)
\oplus\ell^2([j+1,\infty)\cap\bbZ),
\]
we have
\[
A-\bigl(A_{-,j}\oplus A_{+,j+1}\bigr)
=\begin{pmatrix}0&B_j\\B_j^*&0\end{pmatrix}\in\HS,
\]
so this difference is compact.
By Weyl's theorem,
\[
J\cap\sigma_{\rm ess}(A_{-,j})
=
J\cap\sigma_{\rm ess}(A_{+,j+1})
=\varnothing.
\]
Consequently,
\[
\begin{gathered}
\sigma(A_{\pm,j})\cap K\subset\sigma_{\rm disc}(A_{\pm,j}),
\qquad j\in\bbZ,\\
\#\bigl(\Sigma_{\rm edge}(A)\cap K\bigr)
\le\sum_{j=0}^{P-1}\left[
\#\bigl(\sigma(A_{-,j})\cap K\bigr)
+\#\bigl(\sigma(A_{+,j})\cap K\bigr)\right]<\infty.
\end{gathered}
\]

Let $\psi_L$ be a normalized eigenvector of $A_{I_L}$ with eigenvalue $E_L$,
extended by zero to $\ell^2(\bbZ)$, and set
\[
g:=\dist(K,\sigma(A))>0.
\]
Then
\begin{equation}
g
\le \dist(E_L,\sigma(A))
\le \|(A-E_L)\psi_L\|
=\|P_{I_L^c}T_rP_{I_L}\psi_L\|.
\label{eq:endpoint-mass}
\end{equation}
For $b\ge1$, set
\begin{align*}
Q_{L,b}^-&:=P_{[-2PL,-2PL+b-1]\cap\bbZ},\\
Q_{L,b}^+&:=P_{[2PL-b+1,2PL]\cap\bbZ},\\
Q_{L,b}&:=Q_{L,b}^-+Q_{L,b}^+.
\end{align*}
If $x\notin I_L$ and $-2PL+b\le y\le2PL-b$, then
$|x-y|\ge b+1$. Hence Schur's test gives
\[
\|P_{I_L^c}T_r(P_{I_L}-Q_{L,b})\|
\le2\sum_{m\ge b}m^{-r}
\le C_rb^{1-r}.
\]
Choose $b$, independently of $L$, so that $C_rb^{1-r}\le g/2$.
By \eqref{eq:endpoint-mass},
\[
g\le\|P_{I_L^c}T_rQ_{L,b}\psi_L
+P_{I_L^c}T_r(P_{I_L}-Q_{L,b})\psi_L\|
\le\|T_r\|\,\|Q_{L,b}\psi_L\|+g/2.
\]
For sufficiently large $L$,
\[
\|Q_{L,b}^-\psi_L\|^2+\|Q_{L,b}^+\psi_L\|^2
=\|Q_{L,b}\psi_L\|^2\ge\frac{g^2}{4\|T_r\|^2}.
\]
Passing to a subsequence, one of the two boundary layers satisfies the following
lower bound. We treat the right boundary; the left boundary is analogous.
\begin{equation}
\|Q_{L,b}^+\psi_L\|\ge\frac{g}{2\sqrt2\,\|T_r\|}.
\label{eq:right-edge-mass}
\end{equation}
Define
\[
(U_Lf)(n):=f(n+2PL),\qquad
\widetilde\psi_L:=U_L\psi_L,\qquad
\widetilde I_L:=[-4PL,0]\cap\bbZ.
\]
By the $P$-periodicity of $A$,
\[
P_{\widetilde I_L}AP_{\widetilde I_L}\widetilde\psi_L
=E_L\widetilde\psi_L.
\]
Passing to a subsequence, $\widetilde\psi_L\rightharpoonup u$ in
$\ell^2(\bbZ_{\le0})$. Set $Q_b^+:=P_{[1-b,0]\cap\bbZ}$. Then
\[
\|Q_b^+u\|
=\lim_{L\to\infty}\|Q_b^+\widetilde\psi_L\|
=\lim_{L\to\infty}\|Q_{L,b}^+\psi_L\|
\ge\frac{g}{2\sqrt2\,\|T_r\|}>0.
\]
Hence $u\ne0$. For every finitely supported $\varphi\in\ell^2(\bbZ_{\le0})$,
\[
\bigl\langle P_{\widetilde I_L}AP_{\widetilde I_L}\varphi,
\widetilde\psi_L\bigr\rangle
=E_L\langle\varphi,\widetilde\psi_L\rangle,
\]
and
\[
P_{\widetilde I_L}AP_{\widetilde I_L}\varphi
\xrightarrow[L\to\infty]{\ell^2}P_{\bbZ_{\le0}}A\varphi=A_{-,0}\varphi.
\]
Passing to the limit,
\[
\langle A_{-,0}\varphi,u\rangle=E\langle\varphi,u\rangle.
\]
Since finitely supported vectors are dense in $\ell^2(\bbZ_{\le0})$ and
$A_{-,0}$ is bounded and self-adjoint, $A_{-,0}u=Eu$. Hence
$E\in\sigma_{\rm p}(A_{-,0})\subset\Sigma_{\rm edge}(A)$.
\end{proof}

\begin{corollary}\label{cor:uniform-dirichlet-gap}
Under the hypotheses of Lemma~\ref{lem:edge-spectrum}, let $K_0$ be a compact
subset of $J$ satisfying
\[
K_0\cap\Sigma_{\rm edge}(A)=\varnothing.
\]
Then $\sigma(A_{I_L})\cap K_0=\varnothing$ for all sufficiently large $L$.
\end{corollary}

\begin{proof}
If not, choose $L_m\to\infty$ and
$E_m\in\sigma(A_{I_{L_m}})\cap K_0$. Passing to a subsequence,
$E_m\to E\in K_0$. By Lemma~\ref{lem:edge-spectrum},
\[
E\in\Sigma_{\rm edge}(A),
\]
a contradiction.
\end{proof}

\begin{proposition}
\label{prop:finite-template}
For any $\lambda>0$ and $E_*\in\bbR$, there exist a real periodic potential
$w$ with values in $(-\lambda,\lambda)$, a period $P\in\mathbb N$, a number $g>0$, and
$n_*\in\mathbb N$ such that, with
\[
I_n:=[-2Pn,2Pn]\cap\bbZ,
\qquad
\ell_n:=|I_n|=4Pn+1,
\]
one has
\begin{equation}
\dist\!\left(E_*,\sigma((T_r+w)_{I_n+s})\right)\ge8g,
\qquad
n\ge n_*,
\quad
s\in P\bbZ.
\label{eq:finite-template-gap}
\end{equation}
\end{proposition}

\begin{proof}
Since $\mathcal FT_r\mathcal F^{-1}=M_{f_r}$, \eqref{eq:symbol-monotone} yields
\[
\sigma(T_r)=f_r([-\pi,\pi])=[f_r(\pi),f_r(0)].
\]
If $E_*\notin\sigma(T_r)$, take $w=0$. For every finite interval
$I\subset\bbZ$ and $u\in\ell^2(I)$,
\[
f_r(\pi)\|u\|^2
\le\langle u,(T_r)_Iu\rangle
\le f_r(0)\|u\|^2.
\]
Thus \eqref{eq:finite-template-gap} holds for sufficiently small $g>0$.

If $E_*\in\{f_r(\pi),f_r(0)\}$, choose $\kappa$ with
$0<|\kappa|<\lambda/4$, taking $\kappa<0$ when $E_*=f_r(0)$ and
$\kappa>0$ when $E_*=f_r(\pi)$. Then
\[
\dist\!\left(E_*,\sigma(T_r+\kappa I)\right)=|\kappa|>0.
\]
Thus \eqref{eq:finite-template-gap} follows with $w\equiv\kappa$.

If $E_*\in\sigma(T_r)^\circ$, choose a compact interval
$K\Subset(f_r(\pi),f_r(0))$ with $E_*\in K^\circ$. Apply
Theorem~\ref{thm:periodic-gap-input} on $K$, obtaining $\eps_0>0$ and $P_0$,
and choose
\[
0<\eps<\min\{\lambda/8,\eps_0\}.
\]
For sufficiently large $P>P_0$, choose $k\in\{0,\ldots,P\}$ such that
\[
e=f_r(k\pi/P)\in K,
\qquad
|e-E_*|<\lambda/8.
\]
Then
\[
A=T_r+\eps\one_{\{n\equiv0\ (\mathrm{mod}\ P)\}}
\]
has a spectral gap $(e,e+\delta)$ by
Theorem~\ref{thm:periodic-gap-input}. We may assume
\[
0<\delta<\lambda/8.
\]
By Lemma~\ref{lem:edge-spectrum}, choose $c$ and a compact interval $K_0$ such
that
\[
c\in K_0^\circ,
\qquad
K_0\Subset(e,e+\delta)\setminus\Sigma_{\rm edge}(A).
\]
By Corollary~\ref{cor:uniform-dirichlet-gap},
\[
\sigma(A_{I_n})\cap K_0=\varnothing
\]
for all sufficiently large $n$.

Set
\[
\kappa:=E_*-c,\qquad
w(n):=\kappa+\eps\one_{\{n\equiv0\ (\mathrm{mod}\ P)\}}.
\]
Since
\[
|\kappa|\le|E_*-e|+|e-c|<\lambda/4,
\qquad\eps<\lambda/8,
\]
we have $w(n)\in(-\lambda,\lambda)$ for every $n\in\bbZ$. Choose $g>0$ so small that
\[
[E_*-8g,E_*+8g]\subset K_0+\kappa.
\]
Since $w$ is $P$-periodic,
\[
\sigma\bigl((T_r+w)_{I_n+s}\bigr)
=
\sigma\bigl((T_r+w)_{I_n}\bigr),
\qquad
s\in P\bbZ.
\]
Thus \eqref{eq:finite-template-gap} holds for all sufficiently large $n$.
\end{proof}

Write $H:=H_{\lambda,\omega}$. Fix $\lambda>0$ and $E_*\in\bbR$, and choose $w,P,g$ as in
Proposition~\ref{prop:finite-template}. Since
$w(j)/\lambda\in(-1,1)$ for each $j\in\{0,\dots,P-1\}$, the set
\[
\{t\in(-1,1):|\lambda t-w(j)|<g\}
\]
is a nonempty open interval. Choose a nonzero
real-valued $b_j\in C_c^\infty((-1,1))$ and set $\chi_j:=b_j^2$ so that
\begin{equation}
0\le\chi_j\le\frac12,
\qquad
\supp\chi_j
\subset
\{t\in(-1,1):|\lambda t-w(j)|<g\}.
\label{eq:marker-cutoff}
\end{equation}
Let $(\Omega_0,\mathcal F_0,\mathbb P_0)$ be the original probability space.
Enlarge the probability space to
\[
\widetilde\Omega=\Omega_0\times[0,1]^{\bbZ},
\qquad
\widetilde{\mathcal F}
=
\mathcal F_0\otimes\mathcal B([0,1])^{\otimes\bbZ},
\qquad
\widetilde{\mathbb P}
=
\mathbb P_0\otimes\operatorname{Leb}_{[0,1]}^{\otimes\bbZ}.
\]
Henceforth we work on this enlarged probability space and, for notational
simplicity, write $\mathbb P:=\widetilde{\mathbb P}$ and $\mathbb E$ for the
corresponding expectation. Writing $U_n(\omega,u)=u_n$, the variables
$U_n\sim\operatorname{Unif}[0,1]$ are i.i.d. and independent of the original
potential. Set
\[
\eta_n=\one_{\{U_n<\chi_{n\bmod P}(\omega_n)\}}.
\]
Then
\[
\mathbb P(\eta_n=1\mid\omega_n)=\chi_{n\bmod P}(\omega_n),
\]
and the pairs $(\omega_n,\eta_n)$ are independent over $n$.

\begin{lemma}\label{lem:marked-law}
Conditioned on $\eta=(\eta_n)_{n\in\bbZ}$, the variables $\omega_n$
remain independent, and
\[
\mathbb P(\omega_n\in dt\mid\eta)
=p_{n\bmod P,\eta_n}(t)\,dt,
\qquad n\in\bbZ,
\]
where, for $j=0,\dots,P-1$,
\begin{align}
p_{j,1}(t)
&=
\frac{\chi_j(t)\one_{(-1,1)}(t)}
     {\displaystyle\int_{-1}^1\chi_j(s)\,ds},
\label{eq:marked-density-1}\\
p_{j,0}(t)
&=
\frac{(1-\chi_j(t))\one_{(-1,1)}(t)}
     {\displaystyle\int_{-1}^1(1-\chi_j(s))\,ds}.
\label{eq:marked-density-0}
\end{align}
The finite family $\{p_{j,a}:0\le j<P,\ a\in\{0,1\}\}$ satisfies
\eqref{eq:mixed-info} with a uniform mixed-information bound. Moreover, if
$B=B(\eta)\subset\bbZ$ is $\sigma(\eta)$-measurable, then
\[
\mathcal L\!\left(
(\omega_n)_{n\in B^c}
\,\middle|\,
\eta,(\omega_n)_{n\in B}
\right)
=
\bigotimes_{n\in B^c}p_{n\bmod P,\eta_n}(t_n)\,dt_n.
\]
\end{lemma}

\begin{proof}
For $n\in\bbZ$, let $j=n\bmod P$. The density of $\omega_n$ is
$\frac12\one_{(-1,1)}$. Since
\[
\mathbb P(\eta_n=1\mid\omega_n=t)=\chi_j(t),
\qquad
\mathbb P(\eta_n=0\mid\omega_n=t)=1-\chi_j(t),
\]
we have
\[
\mathbb P(\omega_n\in dt,\eta_n=1)
=\frac12\chi_j(t)\one_{(-1,1)}(t)\,dt,
\]
and
\[
\mathbb P(\omega_n\in dt,\eta_n=0)
=\frac12(1-\chi_j(t))\one_{(-1,1)}(t)\,dt.
\]
Therefore, since $\chi_j\not\equiv0$,
\begin{equation}
q_j:=\mathbb P(\eta_n=1)
=\frac12\int_{-1}^1\chi_j(s)\,ds>0,
\qquad n\equiv j\pmod P,
\label{eq:mark-probability}
\end{equation}
and
\[
\mathbb P(\eta_n=0)=\frac12\int_{-1}^1(1-\chi_j(s))\,ds.
\]
Thus
\begin{align*}
\mathbb P(\omega_n\in dt\mid\eta_n=1)
&=\frac{\chi_j(t)\one_{(-1,1)}(t)}
        {\int_{-1}^1\chi_j(s)\,ds}\,dt
=p_{j,1}(t)\,dt,\\
\mathbb P(\omega_n\in dt\mid\eta_n=0)
&=\frac{(1-\chi_j(t))\one_{(-1,1)}(t)}
        {\int_{-1}^1(1-\chi_j(s))\,ds}\,dt
=p_{j,0}(t)\,dt.
\end{align*}
Since the pairs $(\omega_m,\eta_m)$, $m\in\bbZ$, are independent, for every $n$,
\[
\mathbb P(\omega_n\in dt\mid\eta)
=
p_{n\bmod P,\eta_n}(t)\,dt.
\]

Since $\chi_j=b_j^2$ and $\supp b_j\Subset(-1,1)$,
\[
p_{j,1}(t)
=
\left(\frac{b_j(t)}{\|b_j\|_{L^2(-1,1)}}\right)^2.
\]
For $p_{j,0}$, the function $1-\chi_j$ is smooth on $[-1,1]$ and
\[
\frac12\le1-\chi_j(t)\le1,
\qquad t\in[-1,1].
\]
Lemma~\ref{lem:density} gives finite mixed information for both
$p_{j,1}$ and $p_{j,0}$.
Since the family of densities $p_{j,a}$ is finite, the mixed-information bound
is uniform in $j$ and $a$.

Since $B$ is $\sigma(\eta)$-measurable, conditional independence given $\eta$ yields
\[
\mathcal L\!\left(
(\omega_n)_{n\in B^c}
\,\middle|\,
\eta,(\omega_n)_{n\in B}
\right)
=
\bigotimes_{n\in B^c}p_{n\bmod P,\eta_n}(t_n)\,dt_n.
\]
\end{proof}

We next select barrier blocks. Recall from \eqref{eq:mark-probability} that
$q_j>0$, and set
\[
q:=\min_{0\le j<P}q_j>0.
\]
Choose an integer $a>4$ so large that
\begin{equation}
a q^{4P}>1.
\label{eq:a-choice}
\end{equation}
Recall $I_n=[-2Pn,2Pn]\cap\bbZ$ and $\ell_n=|I_n|=4Pn+1$.

\begin{lemma}\label{lem:barrier-selection}
There exists an almost surely finite, $\sigma(\eta)$-measurable integer $N_0$
such that, for every $n\ge N_0$, one can select $\sigma(\eta)$-measurable
intervals $B_n^\pm$ satisfying
\begin{gather*}
B_n^\pm=I_n+s_n^\pm,\qquad s_n^\pm\in P\bbZ,\\
B_n^+\subset[a^n,2a^n],\qquad B_n^-\subset[-2a^n,-a^n],
\end{gather*}
and $\eta_m=1$ for every $m\in B_n^\pm$. Then
\begin{align}
\max_{m\in B_n^\pm}|\lambda\omega_m-w(m)|&<g,
\label{eq:barrier-pointwise}\\*
\dist(E_*,\sigma(H_{B_n^\pm}))&\ge7g.
\label{eq:single-black-gap}
\end{align}
\end{lemma}

\begin{proof}
For each sufficiently large $n$, choose deterministic families
$\mathcal I_n^\pm$ of pairwise disjoint translates $I_n+s$, $s\in P\bbZ$,
contained in $[a^n,2a^n]$ and $[-2a^n,-a^n]$, respectively.
Since $\ell_n\asymp n$, there is
$c>0$ such that
\[
M_n^\pm:=|\mathcal I_n^\pm|\ge \frac{ca^n}{n}
\]
for all sufficiently large $n$. For $I\in\mathcal I_n^\pm$, set
\[
S_I:=\bigcap_{m\in I}\{\eta_m=1\}.
\]
Then
\[
\mathbb P(S_I)
=\prod_{m\in I}\mathbb P(\eta_m=1)
\ge q^{\ell_n}.
\]
Define
\[
F_n^\pm:=\bigcap_{I\in\mathcal I_n^\pm}S_I^c.
\]
The events $S_I$, $I\in\mathcal I_n^\pm$, are independent.
Thus, for sufficiently large $n_0\ge n_*$,
\begin{equation}
\begin{aligned}
\sum_{n\ge n_0}\mathbb P(F_n^\pm)
&\le\sum_{n\ge n_0}(1-q^{\ell_n})^{M_n^\pm}
\le\sum_{n\ge n_0}\exp(-M_n^\pm q^{\ell_n})\\
&\le\sum_{n\ge n_0}\exp\!\left(-\frac{cq}{n}(aq^{4P})^n\right)<\infty.
\end{aligned}
\label{eq:barrier-failure}
\end{equation}
Since the events $F_n^\pm$ are $\sigma(\eta)$-measurable, the Borel--Cantelli
lemma gives an almost surely finite, $\sigma(\eta)$-measurable integer
$N_0\ge n_0$ such that neither $F_n^+$ nor $F_n^-$ occurs for any $n\ge N_0$.
For such $n$, let $B_n^\pm$ be the leftmost interval
$I\in\mathcal I_n^\pm$ for which $S_I$ occurs. This choice is
$\sigma(\eta)$-measurable. For every $m\in B_n^\pm$, we have $\eta_m=1$, so
$\chi_{m\bmod P}(\omega_m)>0$. By \eqref{eq:marker-cutoff} and the
$P$-periodicity of $w$, this proves \eqref{eq:barrier-pointwise}.
Set $H_{B_n^\pm}^{(w)}:=(T_r+w)_{B_n^\pm}$.
By Proposition~\ref{prop:finite-template} and $N_0\ge n_*$,
$\dist(E_*,\sigma(H_{B_n^\pm}^{(w)}))\ge8g$. By \eqref{eq:barrier-pointwise},
\[
\|H_{B_n^\pm}-H_{B_n^\pm}^{(w)}\|
=\max_{m\in B_n^\pm}|\lambda\omega_m-w(m)|<g.
\]
Hence, by the standard spectral perturbation bound,
\[
\dist(E_*,\sigma(H_{B_n^\pm}))\ge8g-g=7g.\qedhere
\]
\end{proof}

\begin{lemma}\label{lem:black-union-gap}
There exists an almost surely finite, $\sigma(\eta)$-measurable integer $N$ such that
$B_n^\pm$ are defined for all $n\ge N$ and, with
\[
B:=\bigcup_{n\ge N}(B_n^+\cup B_n^-),
\]
one has
\begin{equation}
\dist(E_*,\sigma(H_B))\ge6g.
\label{eq:black-union-gap}
\end{equation}
Consequently, $R_B(z)=(H_B-z)^{-1}$ is uniformly bounded for
$|z-E_*|\le3g$ and satisfies all conclusions of
Proposition~\ref{prop:uniform-inverse} there.
\end{lemma}

\begin{proof}
Let $N_0$ and $B_n^\pm$ be given by Lemma~\ref{lem:barrier-selection}.
For $M\ge N_0$, set
\[
B^{(M)}:=\bigcup_{n\ge M}(B_n^+\cup B_n^-).
\]
Then
\[
|B_n^\pm|=4Pn+1\le C(1+n).
\]
Moreover, any two distinct blocks $B_n^\varepsilon$ and
$B_m^{\varepsilon'}$, with $\varepsilon,\varepsilon'\in\{+,-\}$, satisfy
\begin{equation}
\dist(B_n^\varepsilon,B_m^{\varepsilon'})
\ge \left(1-\frac2a\right)a^{\max\{n,m\}}.
\label{eq:black-block-separation}
\end{equation}
Thus,
\begin{align}
\|T_{B_n^\varepsilon B_m^{\varepsilon'}}\|
&\le\|T_{B_n^\varepsilon B_m^{\varepsilon'}}\|_{\HS}\notag\\
&=\left(\sum_{x\in B_n^\varepsilon}\sum_{y\in B_m^{\varepsilon'}}|x-y|^{-2r}\right)^{1/2}\notag\\
&\le\sqrt{|B_n^\varepsilon|\,|B_m^{\varepsilon'}|}\,\dist(B_n^\varepsilon,B_m^{\varepsilon'})^{-r}\notag\\*
&\le C\sqrt{(1+n)(1+m)}\,a^{-r\max\{n,m\}}.
\label{eq:black-block-coupling}
\end{align}

Set
\[
H_{B^{(M)}}^{(0)}
:=
\bigoplus_{n\ge M}\bigl(H_{B_n^+}\oplus H_{B_n^-}\bigr).
\]
By \eqref{eq:single-black-gap},
\[
\dist(E_*,\sigma(H_{B^{(M)}}^{(0)}))\ge7g.
\]
Set
\[
K_M:=H_{B^{(M)}}-H_{B^{(M)}}^{(0)}.
\]
By \eqref{eq:black-block-coupling}, for fixed
$n\ge M$,
\begin{align*}
\sum_{\substack{m\ge M,\ \varepsilon'\in\{+,-\}\\
(m,\varepsilon')\ne(n,\varepsilon)}}
\|T_{B_n^\varepsilon B_m^{\varepsilon'}}\|
&\le
C\sum_{M\le m\le n}\sqrt{(1+n)(1+m)}\,a^{-rn}\\*
&\quad+
C\sum_{m>n}\sqrt{(1+n)(1+m)}\,a^{-rm}\\*
&\le C(1+n)^2a^{-rn}.
\end{align*}
The block-column sums satisfy the same bound. By the block Schur test,
\[
\|K_M\|
\le C\sup_{n\ge M}(1+n)^2a^{-rn}
\longrightarrow0
\qquad(M\to\infty).
\]
Choose $M_1$ sufficiently large that
\[
C\sup_{k\ge M}(1+k)^2a^{-rk}<g,
\qquad M\ge M_1,
\]
and set $N:=\max\{N_0,M_1\}$. With
$B=B^{(N)}$ and $K=K_N$,
\[
H_B=H_B^{(0)}+K,
\qquad
\|K\|<g.
\]
By the standard spectral perturbation bound,
\[
\dist(E_*,\sigma(H_B))\ge7g-\|K\|\ge6g,
\]
which is \eqref{eq:black-union-gap}.

If $|z-E_*|\le3g$, then
\[
\dist(z,\sigma(H_B))
\ge\dist(E_*,\sigma(H_B))-|z-E_*|
\ge3g.
\]
Hence $\|R_B(z)\|\le(3g)^{-1}$, and
Proposition~\ref{prop:uniform-inverse} applies for $|z-E_*|\le3g$.
\end{proof}

Let
\[
\cF_B:=\sigma\!\left(\eta,(\omega_m)_{m\in B}\right).
\]
Henceforth, conditional on $\cF_B$, the variables $(\omega_m)_{m\in B^c}$ are
independent with densities $p_{m\bmod P,\eta_m}$ by
Lemma~\ref{lem:marked-law}.

Define
\begin{align*}
W_0
&:=\{m\in\bbZ:\max B_N^-<m<\min B_N^+\},\\
W_k
&:=\{m\in\bbZ:\max B_{N+k-1}^+<m<\min B_{N+k}^+\},
\qquad k\ge1,\\
W_{-k}
&:=\{m\in\bbZ:\max B_{N+k}^-<m<\min B_{N+k-1}^-\},
\qquad k\ge1.
\end{align*}
Then
\[
B^c=\bigsqcup_{i\in\bbZ}W_i.
\]
Set
\[
N_i:=|W_i|,\qquad i\in\bbZ.
\]
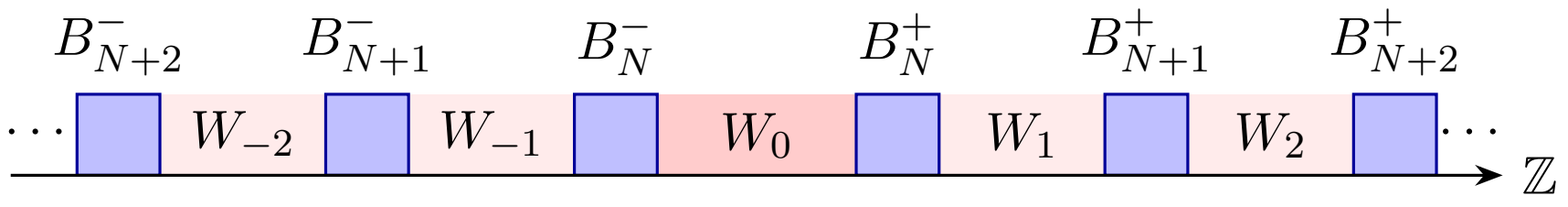
\begin{figure}[H]
\centering
\begin{tikzpicture}[
    x=0.60cm,
    y=0.9cm,
    >=Stealth,
    barrier/.style={draw=blue!60!black,fill=blue!25,line width=0.6pt},
    component/.style={fill=red!8},
    central/.style={fill=red!20}
]
\fill[component] (-7.2,0) rectangle (-5.2,0.65);
\fill[component] (-4.2,0) rectangle (-2.2,0.65);
\fill[central]   (-1.2,0) rectangle ( 1.2,0.65);
\fill[component] ( 2.2,0) rectangle ( 4.2,0.65);
\fill[component] ( 5.2,0) rectangle ( 7.2,0.65);
\draw[barrier] (-8.2,0) rectangle (-7.2,0.65);
\draw[barrier] (-5.2,0) rectangle (-4.2,0.65);
\draw[barrier] (-2.2,0) rectangle (-1.2,0.65);
\draw[barrier] ( 1.2,0) rectangle ( 2.2,0.65);
\draw[barrier] ( 4.2,0) rectangle ( 5.2,0.65);
\draw[barrier] ( 7.2,0) rectangle ( 8.2,0.65);
\draw[->,line width=0.6pt] (-9,0) -- (9,0) node[right] {$\bbZ$};
\node at (-8.65,0.34) {$\cdots$};
\node at ( 8.65,0.34) {$\cdots$};
\node[above] at (-7.7,0.65) {$B_{N+2}^{-}$};
\node[above] at (-4.7,0.65) {$B_{N+1}^{-}$};
\node[above] at (-1.7,0.65) {$B_N^{-}$};
\node[above] at ( 1.7,0.65) {$B_N^{+}$};
\node[above] at ( 4.7,0.65) {$B_{N+1}^{+}$};
\node[above] at ( 7.7,0.65) {$B_{N+2}^{+}$};
\node at (-6.2,0.32) {$W_{-2}$};
\node at (-3.2,0.32) {$W_{-1}$};
\node at ( 0,0.32) {$W_0$};
\node at ( 3.2,0.32) {$W_1$};
\node at ( 6.2,0.32) {$W_2$};
\end{tikzpicture}
\caption{Barrier blocks and complementary components.}
\label{fig:barriers}
\end{figure}

There exist $\cF_B$-measurable random constants $C_B,c_B\in(0,\infty)$ such that
\begin{align}
N_i&\le C_B a^{|i|},
&
\sup_{x\in W_i}\langle x\rangle&\le C_B a^{|i|},
\label{eq:white-size}\\*
\dist(W_i,W_j)&\ge c_B(1+|i|),
&&|i-j|=1,
\label{eq:white-neighbor}\\*
\dist(W_i,W_j)&\ge c_B a^{\max\{|i|,|j|\}},
&&|i-j|\ge2.
\label{eq:white-far}
\end{align}
These estimates follow directly from the definitions of $W_i$ and the inclusions
$B_n^+\subset[a^n,2a^n]$, $B_n^-\subset[-2a^n,-a^n]$, together with
$|B_n^\pm|=4Pn+1$ and $a>4$.

\section{Schur complement and fractional moments}
\label{sec:schur-factorization}
\label{sec:deletion}

Set
\[
W=B^c=\bigsqcup_{i\in\bbZ}W_i,
\qquad
\ell^2(\bbZ)=\ell^2(B)\oplus\ell^2(W).
\]
For $z\in\bbC\setminus\bbR$, set
\begin{equation}
R_B(z):=(H_B-z)^{-1},\qquad
S(z):=H_W-z-T_{WB}R_B(z)T_{BW}.
\label{eq:BW-block-matrix}
\end{equation}
By the Schur complement formula,
\begin{equation}
P_W(H-z)^{-1}P_W=S(z)^{-1}.
\label{eq:schur-inverse}
\end{equation}
With respect to $\ell^2(W)=\bigoplus_i\ell^2(W_i)$,
\begin{align}
S(z)&=D(z)+V(z),\qquad D(z)=\bigoplus_iD_i(z),
\label{eq:schur}\\
D_i(z)&=Q_i^{(0)}(z)+\lambda\diag(\omega|_{W_i}),
\qquad
Q_i^{(0)}(z):=T_{W_iW_i}-z-T_{W_iB}R_B(z)T_{BW_i},
\label{eq:D-blocks}\\
V_{ij}(z)&=T_{W_iW_j}-T_{W_iB}R_B(z)T_{BW_j},
\quad i\ne j,
\qquad
V_{ii}(z)=0.
\label{eq:V-blocks}
\end{align}
The operators $Q_i^{(0)}(z)$ and $V_{ij}(z)$ are $\cF_B$-measurable.

For $|z-E_*|\le3g$, Lemma~\ref{lem:black-union-gap} gives
\[
\dist(z,\sigma(H_B))\ge3g.
\]
Propositions~\ref{prop:uniform-inverse} and~\ref{prop:rectangular-self-energy} yield, for $i\ne j$,
\begin{equation}
\gamma(V_{ij}(z))\le C_Bd_{ij}^{-\alpha},
\qquad
\|V_{ij}(z)\|_{\TC}
\le C_B\sqrt{N_iN_j}\,d_{ij}^{-r},
\qquad
d_{ij}:=\dist(W_i,W_j).
\label{eq:Vij-basic}
\end{equation}

\begin{lemma}\label{lem:taylor}
There exists $h>0$, uniform over all choices of $B$ and $\omega|_B$ satisfying
$\dist(E_*,\sigma(H_B))\ge6g$, with the following property.
For each such choice, there exist $\cF_B$-measurable operators $A_{ij,\nu}$,
$i\ne j$, $\nu\ge0$, such that, for $|z-E_*|<h$,
\begin{equation}
V_{ij}(z)
=
\sum_{\nu\ge0}A_{ij,\nu}
\left(\frac{z-E_*}{h}\right)^\nu,
\qquad
A_{ji,\nu}=A_{ij,\nu}^*,
\label{eq:V-taylor}
\end{equation}
and
\begin{equation}
\gamma(A_{ij,\nu})\le C_B2^{-\nu}d_{ij}^{-\alpha},
\qquad
\|A_{ij,\nu}\|_{\TC}
\le C_B2^{-\nu}\sqrt{N_iN_j}\,d_{ij}^{-r}.
\label{eq:taylor-coeff-bounds}
\end{equation}
\end{lemma}

\begin{proof}
Set
\[
R_0:=(H_B-E_*)^{-1}.
\]
By Proposition~\ref{prop:uniform-inverse},
\begin{equation}
\|R_0\|_{J_r}+\|R_0\|_{\mathcal C}+g_\alpha(R_0)\le C_0.
\label{eq:R0-basic}
\end{equation}
Equations \eqref{eq:J-product}, \eqref{eq:C-J} and
Lemma~\ref{lem:crosscut-product} imply, after increasing $C_0$,
\begin{equation}
\|R_0^k\|_{J_r}+\|R_0^k\|_{\mathcal C}+g_\alpha(R_0^k)
\le C_0^k,
\qquad k\ge1.
\label{eq:R0-powers}
\end{equation}
Applying Lemma~\ref{lem:crosscut-product} twice gives
\begin{align}
g_\alpha(T_rR_0^k)^{p_0}
&\le C_\alpha\Bigl(
\|T_r\|_{\mathcal C}^{p_0}g_\alpha(R_0^k)^{p_0}
+g_\alpha(T_r)^{p_0}\|R_0^k\|_{\mathcal C}^{p_0}
\Bigr)
\le C C_0^{kp_0},
\label{eq:gamma-one-product}\\
g_\alpha(T_rR_0^kT_r)^{p_0}
&\le C_\alpha\Bigl(
\|T_rR_0^k\|_{\mathcal C}^{p_0}g_\alpha(T_r)^{p_0}
+g_\alpha(T_rR_0^k)^{p_0}\|T_r\|_{\mathcal C}^{p_0}
\Bigr)
\le C C_0^{kp_0}.
\label{eq:gamma-two-products}
\end{align}
By \eqref{eq:gamma-compression} and the definition of $g_\alpha$,
\begin{equation}
\begin{aligned}
\gamma(T_{W_iB}R_0^kT_{BW_j})
&=\gamma(P_{W_i}T_rR_0^kT_rP_{W_j})\\
&\le d_{ij}^{-\alpha}g_\alpha(T_rR_0^kT_r)
\le CC_0^k d_{ij}^{-\alpha}.
\end{aligned}
\label{eq:R0-power-gamma}
\end{equation}
By \eqref{eq:rectangular-self-energy} and \eqref{eq:R0-powers},
\begin{equation}
\|T_{W_iB}R_0^kT_{BW_j}\|_{\TC}
\le CC_0^k\sqrt{N_iN_j}\,d_{ij}^{-r}.
\label{eq:R0-power-trace}
\end{equation}
Choose
\begin{equation}
0<h\le\min\{3g,(2C_0)^{-1}\}.
\label{eq:h-choice}
\end{equation}
Since $\|R_0\|\le(6g)^{-1}$,
\[
\|(z-E_*)R_0\|\le\frac{|z-E_*|}{6g}<1.
\]
The Neumann series gives
\begin{equation}
R_B(z)
=\bigl[I-(z-E_*)R_0\bigr]^{-1}R_0
=\sum_{\nu\ge0}(z-E_*)^\nu R_0^{\nu+1}.
\label{eq:RB-neumann}
\end{equation}
Consequently,
\[
V_{ij}(z)=V_{ij}(E_*)-
\sum_{\nu\ge1}h^\nu T_{W_iB}R_0^{\nu+1}T_{BW_j}
\left(\frac{z-E_*}{h}\right)^\nu.
\]
It follows that \eqref{eq:V-taylor} holds with
\begin{equation}
A_{ij,0}:=V_{ij}(E_*),
\qquad
A_{ij,\nu}:=-h^\nu T_{W_iB}R_0^{\nu+1}T_{BW_j},
\quad\nu\ge1.
\label{eq:Aij-definition}
\end{equation}
Equations \eqref{eq:R0-power-gamma}, \eqref{eq:R0-power-trace} and
$hC_0\le1/2$ imply, for $\nu\ge1$,
\[
\begin{aligned}
\gamma(A_{ij,\nu})
&\le C(hC_0)^\nu d_{ij}^{-\alpha}
\le C2^{-\nu}d_{ij}^{-\alpha},\\
\|A_{ij,\nu}\|_{\TC}
&\le C(hC_0)^\nu\sqrt{N_iN_j}\,d_{ij}^{-r}
\le C2^{-\nu}\sqrt{N_iN_j}\,d_{ij}^{-r}.
\end{aligned}
\]
The case $\nu=0$ is \eqref{eq:Vij-basic}. Finally,
\[
R_0^*=R_0
\quad\Longrightarrow\quad
A_{ij,\nu}^*=A_{ji,\nu}.
\]
\end{proof}

For each $i<j$ and $\nu\ge0$, choose an admissible factorization of
$A_{ij,\nu}$ in the sense of Definition~\ref{def:gamma}, with $\cF_B$-measurable factors,
\[
A_{ij,\nu}
=L_{i;\{i,j\},\nu}^*L_{j;\{i,j\},\nu},
\]
where, for an auxiliary Hilbert space $\cK_{\{i,j\},\nu}$,
\[
L_{i;\{i,j\},\nu}:\ell^2(W_i)\to\cK_{\{i,j\},\nu},
\qquad
L_{j;\{i,j\},\nu}:\ell^2(W_j)\to\cK_{\{i,j\},\nu}.
\]
For $|i-j|=1$, choose the factors by Lemma~\ref{lem:measurable-balanced}(i).
By \eqref{eq:taylor-coeff-bounds} and \eqref{eq:white-neighbor},
\begin{equation}
\Phi(L_{i;\{i,j\},\nu})=\Phi(L_{j;\{i,j\},\nu})
\le C_B2^{-\nu}d_{ij}^{-\alpha}
\le C_B2^{-\nu}(1+|i|)^{-\alpha}.
\label{eq:near-factor}
\end{equation}
For $|i-j|\ge2$, choose the factors by Lemma~\ref{lem:measurable-balanced}(ii).
By \eqref{eq:taylor-coeff-bounds},
\begin{equation}
\begin{aligned}
\Phi(L_{i;\{i,j\},\nu})&=\Phi(L_{j;\{i,j\},\nu})\\
&\le(N_iN_j)^{1/4}\|A_{ij,\nu}\|_{\TC}
\le C_B2^{-\nu}(N_iN_j)^{3/4}d_{ij}^{-r}.
\end{aligned}
\label{eq:far-factor}
\end{equation}
By \eqref{eq:white-size}, \eqref{eq:white-far}, and \eqref{eq:far-factor},
\begin{align}
\Phi(L_{i;\{i,j\},\nu})
&=\Phi(L_{j;\{i,j\},\nu})\notag\\
&\le C_B2^{-\nu}(N_iN_j)^{3/4}d_{ij}^{-r}\notag\\
&\le C_B2^{-\nu}a^{\frac34(|i|+|j|)-r\max\{|i|,|j|\}}\notag\\
&=C_B2^{-\nu}a^{-\alpha\min\{|i|,|j|\}-(r-\frac34)\bigl||i|-|j|\bigr|}.
\label{eq:far-factor-geometric}
\end{align}

Fix $0\le\tau<\alpha$ and choose
\begin{equation}
\tau+\frac12<\beta<r-1.
\label{eq:beta-choice}
\end{equation}
Set
\begin{equation}
c_{ij}=
\begin{cases}
1,&|i-j|=1,\\
a^{-\beta\,\bigl||i|-|j|\bigr|},&|i-j|\ge2.
\end{cases}
\label{eq:cij}
\end{equation}
Define
\begin{equation}
\cK_i:=\bigoplus_{\substack{j\ne i\\\nu\ge0}}\cK_{\{i,j\},\nu},
\qquad
F_i:=\bigoplus_{\substack{j\ne i\\\nu\ge0}}
c_{ij}^{-1/2}L_{i;\{i,j\},\nu}:
\ell^2(W_i)\to\cK_i.
\label{eq:Fi}
\end{equation}

\begin{lemma}\label{lem:Fi-decay}
For every $i\in\bbZ$,
\[
F_i\in\HS\bigl(\ell^2(W_i),\cK_i\bigr),
\]
and
\begin{equation}
\Phi(F_i)\le C_B(1+|i|)^{-\alpha}\longrightarrow0
\qquad (|i|\to\infty).
\label{eq:Fi-decay}
\end{equation}
\end{lemma}

\begin{proof}
Let $p_0=2/3$ and
\[
b:=r-\frac34-\beta>\frac14.
\]
For a finite set
$E\subset\{(j,\nu):j\ne i,\ \nu\ge0\}$, define
\[
F_i^{(E)}
:=\bigoplus_{(j,\nu)\in E}
c_{ij}^{-1/2}L_{i;\{i,j\},\nu}.
\]
Then
\[
\Phi(F_i^{(E)})^{p_0}
\le
\sum_{(j,\nu)\in E}
c_{ij}^{-p_0}\Phi(L_{i;\{i,j\},\nu})^{p_0}.
\]
Grouping the terms by $k=|j|$ and using
\eqref{eq:near-factor}, \eqref{eq:far-factor-geometric}, and \eqref{eq:cij}, we obtain
\begin{align}
\Phi(F_i^{(E)})^{p_0}
&\le\sum_{\substack{(j,\nu)\in E\\|i-j|=1}}\Phi(L_{i;\{i,j\},\nu})^{p_0}
+\sum_{\substack{(j,\nu)\in E\\|i-j|\ge2,\ |j|\le |i|}}c_{ij}^{-p_0}\Phi(L_{i;\{i,j\},\nu})^{p_0}\notag\\*
&\quad+\sum_{\substack{(j,\nu)\in E\\|i-j|\ge2,\ |j|>|i|}}c_{ij}^{-p_0}\Phi(L_{i;\{i,j\},\nu})^{p_0}\notag\\*
&\le C_B(1+|i|)^{-p_0\alpha}
+C_B\sum_{k=0}^{|i|}a^{-p_0[\alpha k+b(|i|-k)]}
+C_B\sum_{k>|i|}a^{-p_0[\alpha |i|+b(k-|i|)]}
\notag\\*
&\le C_B(1+|i|)^{-p_0\alpha}
+C_B(1+|i|)a^{-p_0\min\{\alpha,b\}|i|}
+C_Ba^{-p_0\alpha |i|}
\notag\\*
&\le C_B(1+|i|)^{-p_0\alpha}.
\label{eq:Fi-p0-estimate}
\end{align}
Now let
\[
E_R:=\{(j,\nu):j\ne i,\ |j|\le R,\ 0\le\nu\le R\}.
\]
For $R'>R$,
\[
\Phi\!\left(F_i^{(E_{R'})}-F_i^{(E_R)}\right)^{p_0}
\le
\sum_{(j,\nu)\in E_{R'}\setminus E_R}
c_{ij}^{-p_0}\Phi(L_{i;\{i,j\},\nu})^{p_0}
\longrightarrow0.
\]
By \eqref{eq:phi-hs},
\[
\left\|F_i^{(E_{R'})}-F_i^{(E_R)}\right\|_{\HS}
\longrightarrow0.
\]
Thus $F_i^{(E_R)}$ converges to $F_i$ in $\HS(\ell^2(W_i),\cK_i)$.
Applying \eqref{eq:Fi-p0-estimate} with $E=E_R$ and letting $R\to\infty$ yields \eqref{eq:Fi-decay}.
\end{proof}

By \eqref{eq:phi-hs} and \eqref{eq:Fi-decay},
\[
\operatorname{rank}F_i\le N_i<\infty,
\qquad
\|F_i\|^2\le\|F_i\|_{\HS}^2\le\Phi(F_i),
\qquad\sup_i\|F_i\|<\infty.
\]
Hence
\[
F:=\bigoplus_iF_i\in
\mathcal B\!\left(\ell^2(W),\bigoplus_i\cK_i\right).
\]

Set
\[
\cK:=\bigoplus_{i\in\bbZ}\cK_i
=\bigoplus_{\substack{i<j\\\nu\ge0}}
\left(\cK_{\{i,j\},\nu}\oplus\cK_{\{i,j\},\nu}\right).
\]
Define
\[
t(z):=\frac{z-E_*}{h},
\qquad
J_{ij,\nu}(z)
:=c_{ij}t(z)^\nu
\begin{pmatrix}
0&I_{\cK_{\{i,j\},\nu}}\\
I_{\cK_{\{i,j\},\nu}}&0
\end{pmatrix},
\]
and set
\begin{equation}
J(z):=
\bigoplus_{\substack{i<j\\\nu\ge0}}J_{ij,\nu}(z)
\quad\text{on }\cK.
\label{eq:J-definition}
\end{equation}
Then
\[
J(z)=J(\bar z)^*,
\qquad
(F^*J(z)F)_{ij}=V_{ij}(z),
\qquad
(F^*J(z)F)_{ii}=0,
\]
so
\begin{equation}
V(z)=F^*J(z)F.
\label{eq:V-FJF}
\end{equation}
For $|z-E_*|\le h/2$,
\begin{equation}
\|J_{ij}(z)\|\le c_{ij}.
\label{eq:Jij-basic}
\end{equation}
For $\theta\ge0$, set
\begin{equation}
X_\theta
:=
\left\{q=(q_i):q_i\in\cK_i,\ 
\|q\|_{X_\theta}^2:=\sum_i a^{2\theta|i|}\|q_i\|^2<\infty\right\}.
\label{eq:Xtheta}
\end{equation}
\begin{lemma}\label{lem:J-weighted}
Let $0\le\theta<\beta$. For every $s>0$, there exists
$C_{s,\theta}<\infty$ such that, for all $|z-E_*|\le h/2$,
\begin{equation}
\sup_i\sum_{j\ne i}\|J_{ij}(z)\|^s a^{s\theta(|i|-|j|)}
\le C_{s,\theta}.
\label{eq:J-sum}
\end{equation}
Moreover, there exists $C_\theta<\infty$ such that, whenever
$|z-E_*|,|w-E_*|\le h/2$,
\begin{equation}
\|J(z)\|_{X_\theta\to X_\theta}\le C_\theta,
\label{eq:J-bounded}
\end{equation}
and
\begin{equation}
\|J(z)-J(w)\|_{X_\theta\to X_\theta}
\le C_\theta\frac{|z-w|}{h}.
\label{eq:J-continuity}
\end{equation}
\end{lemma}

\begin{proof}
By \eqref{eq:Jij-basic} and \eqref{eq:cij}, for every $s>0$,
\begin{align*}
\sup_i\sum_{j\ne i}\|J_{ij}(z)\|^s a^{s\theta(|i|-|j|)}
&\le Ca^{s\theta}
+C\sup_i\sum_{|i-j|>1}c_{ij}^s a^{s\theta(|i|-|j|)}
\\
&=Ca^{s\theta}
+C\sup_i\sum_{|i-j|>1}
a^{-s\beta\,\bigl||i|-|j|\bigr|}a^{s\theta(|i|-|j|)}
\\
&\le Ca^{s\theta}
+C\sup_i\sum_{|i-j|>1}
a^{-s(\beta-\theta)\,\bigl||i|-|j|\bigr|}
<\infty.
\end{align*}
The same argument gives
\begin{equation}
\sup_j\sum_{i\ne j}\|J_{ij}(z)\|a^{\theta(|i|-|j|)}<\infty.
\label{eq:J-column-sum}
\end{equation}
By \eqref{eq:J-sum} with $s=1$, \eqref{eq:J-column-sum}, and the weighted Schur test,
\[
\|J(z)\|_{X_\theta\to X_\theta}\le C_\theta.
\]
Moreover, if $|z-E_*|,|w-E_*|\le h/2$, then
$|t(z)|,|t(w)|\le1/2$. Hence, for $\nu\ge1$,
\[
|t(z)^\nu-t(w)^\nu|
\le |t(z)-t(w)|\sum_{k=0}^{\nu-1}|t(z)|^k|t(w)|^{\nu-1-k}
\le \nu2^{-(\nu-1)}\frac{|z-w|}{h}.
\]
Then
\[
\|J_{ij}(z)-J_{ij}(w)\|
\le c_{ij}\sum_{\nu\ge1}\nu2^{-(\nu-1)}\frac{|z-w|}{h}
\le Cc_{ij}\frac{|z-w|}{h}.
\]
By Schur's test,
\begin{align*}
&\|J(z)-J(w)\|_{X_\theta\to X_\theta}\\*
&\le\left(\sup_i\sum_{j\ne i}a^{\theta(|i|-|j|)}\|J_{ij}(z)-J_{ij}(w)\|\right)^{1/2}
\times\left(\sup_j\sum_{i\ne j}a^{\theta(|i|-|j|)}\|J_{ij}(z)-J_{ij}(w)\|\right)^{1/2}\\*
&\le C\frac{|z-w|}{h}
\left(\sup_i\sum_{j\ne i}c_{ij}a^{\theta(|i|-|j|)}\right)^{1/2}
\times\left(\sup_j\sum_{i\ne j}c_{ij}a^{\theta(|i|-|j|)}\right)^{1/2}\\*
&\le C_\theta\frac{|z-w|}{h}.
\end{align*}
\end{proof}

Fix
\begin{equation}
0<s<\frac23,
\qquad
I=(E_*-h/4,E_*+h/4),
\qquad
z=E+i\eta,
\quad E\in I,
\quad0<\eta<h/4.
\label{eq:local-z}
\end{equation}
Conditionally on $\cF_B$,
\begin{equation}
\mathbb P_B=\bigotimes_{i\in\bbZ}\mathbb P_{B,i},
\qquad
\mathbb P_{B,i}
=\bigotimes_{n\in W_i}p_{n\bmod P,\eta_n}(t_n)\,dt_n.
\label{eq:conditional-product}
\end{equation}
Throughout the conditional argument in Sections~\ref{sec:schur-factorization}
and~\ref{sec:terminal}, $\mathbb E$ denotes expectation under $\mathbb P_B$.
Let $M_*$ be a uniform mixed-information bound for the density family in Lemma~\ref{lem:marked-law}.
Let
\[
\mathcal G_i:=\sigma(\omega_n:n\in B^c\setminus W_i),
\qquad
\mathbb E_i:=\mathbb E(\,\cdot\mid\mathcal G_i).
\]
For $i\ne j$, similarly define
\[
\mathcal G_{i,j}
:=\sigma\!\left(\omega_n:n\in B^c\setminus(W_i\cup W_j)\right),
\qquad
\mathbb E_{i,j}:=\mathbb E(\,\cdot\mid\mathcal G_{i,j}).
\]
For finite $\Lambda\subset\bbZ$, set
\[
W_\Lambda:=\bigcup_{i\in\Lambda}W_i,
\qquad
\mathcal H_\Lambda:=\ell^2(B\cup W_\Lambda),
\qquad
H^\Lambda:=H_{B\cup W_\Lambda}.
\]
Define
\[
\begin{aligned}
G^\Lambda(z)
&:=P_{W_\Lambda}(H^\Lambda-z)^{-1}P_{W_\Lambda},\\
G_{ij}^\Lambda(z)
&:=P_{W_i}(H^\Lambda-z)^{-1}P_{W_j},
\qquad i,j\in\Lambda,
\end{aligned}
\]
and
\begin{equation}
K_{ij}^\Lambda(z):=F_iG_{ij}^\Lambda(z)F_j^*,
\qquad i,j\in\Lambda,
\label{eq:Kij}
\end{equation}
with $K_{ij}^\Lambda=0$ if $i\notin\Lambda$ or $j\notin\Lambda$.

\begin{lemma}\label{lem:apriori}
For every $i\in\Lambda$,
\begin{equation}
\mathbb E_i\|K_{ii}^\Lambda(z)\|_{\HS}^s
\le b_i,
\qquad
b_i:=C_{s,\lambda,M_*}\Phi(F_i)^s\longrightarrow0.
\label{eq:bi}
\end{equation}
Moreover,
\begin{equation}
\sup_{\Lambda,i,j,z}\mathbb E\|K_{ij}^\Lambda(z)\|_{\HS}^s
\le A_B<\infty.
\label{eq:uniform-K}
\end{equation}
\end{lemma}

\begin{proof}
For fixed $i$, let
\[
X:=B\cup\bigcup_{k\in\Lambda\setminus\{i\}}W_k,
\qquad
H_X:=P_XH^\Lambda P_X.
\]
By the Schur complement with respect to $X$,
\begin{equation}
P_{W_i}(H^\Lambda-z)^{-1}P_{W_i}
=
\bigl(Q_i(z)+\lambda\diag(\omega|_{W_i})\bigr)^{-1},
\label{eq:one-block-schur}
\end{equation}
where
\[
Q_i(z)=T_{W_iW_i}-z-T_{W_iX}(H_X-z)^{-1}T_{XW_i}.
\]
Since $H_X=H_X^*$ and $z=E+i\eta$,
\begin{align*}
\im(H_X-z)^{-1}
&=\frac{(H_X-z)^{-1}-(H_X-\bar z)^{-1}}{2i}\\
&=\frac{z-\bar z}{2i}(H_X-z)^{-1}(H_X-\bar z)^{-1}\\
&=\eta(H_X-z)^{-1}(H_X-\bar z)^{-1}\\
&=\eta\bigl((H_X-E)^2+\eta^2I\bigr)^{-1}\ge0.
\end{align*}
Thus,
\begin{equation}
\im Q_i(z)
=-\eta I-T_{W_iX}\im(H_X-z)^{-1}T_{XW_i}
\le-\eta I<0.
\label{eq:Qi-dissipative}
\end{equation}
Lemma~\ref{lem:multi-site} gives
\[
\mathbb E_i\|K_{ii}^\Lambda(z)\|_{\HS}^s
\le C_{s,\lambda,M_*}\Phi(F_i)^s=b_i.
\]

For $i\ne j$, set
\[
Y=W_i\sqcup W_j,
\qquad
X=(B\cup W_\Lambda)\setminus Y,
\qquad
H_X:=P_XH^\Lambda P_X.
\]
By the Schur complement with respect to $X$,
\begin{equation}
P_Y(H^\Lambda-z)^{-1}P_Y
=
\bigl(Q_{ij}(z)+\lambda\diag(\omega|_Y)\bigr)^{-1},
\label{eq:two-block-schur}
\end{equation}
with
\[
Q_{ij}(z)
=T_{YY}-z-T_{YX}(H_X-z)^{-1}T_{XY},
\]
and
\begin{equation}
\im Q_{ij}(z)
=-\eta I_Y-T_{YX}\im\!\bigl((H_X-z)^{-1}\bigr)T_{XY}
\le-\eta I_Y<0.
\label{eq:Qij-dissipative}
\end{equation}
Let
\[
F_{ij}^{(2)}:=F_i\oplus F_j:
\ell^2(W_i)\oplus\ell^2(W_j)\to\cK_i\oplus\cK_j.
\]
The $(i,j)$ block of
\[
F_{ij}^{(2)}
\bigl(Q_{ij}(z)+\lambda\diag(\omega|_Y)\bigr)^{-1}
(F_{ij}^{(2)})^*
\]
is $K_{ij}^\Lambda(z)$. By Lemma~\ref{lem:multi-site},
\[
\mathbb E_{i,j}\|K_{ij}^\Lambda(z)\|_{\HS}^s
\le C_{s,\lambda,M_*}\Phi(F_{ij}^{(2)})^s.
\]
Since
\[
\Phi(F_{ij}^{(2)})^{p_0}
=\Phi(F_i)^{p_0}+\Phi(F_j)^{p_0},
\]
Lemma~\ref{lem:Fi-decay} yields
\begin{align}
\mathbb E_{i,j}\|K_{ij}^\Lambda(z)\|_{\HS}^s
&\le C_{s,\lambda,M_*}
\left(\Phi(F_i)^{p_0}+\Phi(F_j)^{p_0}\right)^{s/p_0}
\notag\\
&\le C_{s,\lambda,M_*}2^{s/p_0}
\left(\sup_k\Phi(F_k)\right)^s
=:A_B<\infty.
\label{eq:two-block-average}
\end{align}
For $i=j$, the bound follows from \eqref{eq:bi}.
Taking expectations yields \eqref{eq:uniform-K}:
\[
\mathbb E\|K_{ij}^\Lambda(z)\|_{\HS}^s\le A_B,
\qquad i,j\in\Lambda.\qedhere
\]
\end{proof}

\begin{lemma}\label{lem:block-recursion}
Let $\Lambda\subset\bbZ$ be finite, let $i,j\in\Lambda$ with $i\ne j$, and let
$z$ be as in \eqref{eq:local-z}. Then
\begin{equation}
\mathbb E_i\|K_{ij}^\Lambda(z)\|^s
\le
b_i\sum_{k\in\Lambda\setminus\{i\}}\|J_{ik}(z)\|^s
\|K_{kj}^{\Lambda\setminus\{i\}}(z)\|^s.
\label{eq:deletion-recursion-conditional}
\end{equation}
Consequently,
\begin{equation}
\mathbb E\|K_{ij}^\Lambda(z)\|^s
\le
b_i\sum_{k\in\Lambda\setminus\{i\}}\|J_{ik}(z)\|^s
\mathbb E\|K_{kj}^{\Lambda\setminus\{i\}}(z)\|^s.
\label{eq:deletion-recursion}
\end{equation}
\end{lemma}

\begin{proof}
Set
\[
S_\Lambda(z):=H_{W_\Lambda}-z-T_{W_\Lambda B}R_B(z)T_{BW_\Lambda}.
\]
Then $G^\Lambda(z)=S_\Lambda(z)^{-1}$ by the Schur complement formula.
For $i\in\Lambda$, the decomposition
\[
\ell^2(W_\Lambda)=\ell^2(W_i)\oplus\ell^2(W_{\Lambda\setminus\{i\}})
\]
gives
\[
S_\Lambda(z)=
\begin{pmatrix}
D_i(z)&B_i(z)\\
C_i(z)&S_{\Lambda\setminus\{i\}}(z)
\end{pmatrix},
\]
where
\[
B_i(z):=(V_{ik}(z))_{k\in\Lambda\setminus\{i\}},
\qquad
C_i(z):=(V_{ki}(z))_{k\in\Lambda\setminus\{i\}}.
\]
By the block inverse formula, for $i\ne j$,
\begin{equation}
G_{ij}^\Lambda
=-G_{ii}^\Lambda
\sum_{k\in\Lambda\setminus\{i\}}V_{ik}G_{kj}^{\Lambda\setminus\{i\}}.
\label{eq:G-deletion-identity}
\end{equation}
By \eqref{eq:V-FJF} and \eqref{eq:G-deletion-identity}, for $i\ne j$,
\begin{align}
K_{ij}^\Lambda
&=F_iG_{ij}^\Lambda F_j^*\notag\\
&=-\sum_{k\in\Lambda\setminus\{i\}}
\bigl(F_iG_{ii}^\Lambda F_i^*\bigr)
J_{ik}\bigl(F_kG_{kj}^{\Lambda\setminus\{i\}}F_j^*\bigr)\notag\\*
&=-K_{ii}^\Lambda
\sum_{k\in\Lambda\setminus\{i\}}
J_{ik}K_{kj}^{\Lambda\setminus\{i\}}.
\label{eq:deletion-identity}
\end{align}
For $k,j\ne i$,
$K_{kj}^{\Lambda\setminus\{i\}}$ is $\mathcal G_i$-measurable. Hence
\eqref{eq:deletion-identity} gives
\[
\mathbb E_i\|K_{ij}^\Lambda\|^s
\le
\mathbb E_i\|K_{ii}^\Lambda\|^s
\left(\sum_{k\in\Lambda\setminus\{i\}}\|J_{ik}\|\,
\|K_{kj}^{\Lambda\setminus\{i\}}\|\right)^s.
\]
Using $\|A\|\le\|A\|_{\HS}$, \eqref{eq:bi}, and subadditivity for $0<s<1$,
we obtain \eqref{eq:deletion-recursion-conditional}.
Taking expectations gives \eqref{eq:deletion-recursion}.
\end{proof}

\begin{proposition}\label{prop:channel-decay}
For every $0<\theta<\beta$, there exists $C<\infty$, depending on $B$ and $\theta$, such that,
for all finite $\Lambda\subset\bbZ$, $i\in\bbZ$, $E\in I$, and
$0<\eta<h/4$,
\begin{equation}
\mathbb E\|K_{i0}^\Lambda(E+i\eta)\|^s
\le Ca^{-s\theta|i|}.
\label{eq:channel-decay}
\end{equation}
Moreover,
\begin{equation}
\mathbb E\left\|
F_iP_{W_i}(H-E-i\eta)^{-1}P_{W_0}F_0^*
\right\|^s
\le Ca^{-s\theta|i|}.
\label{eq:channel-decay-infinite}
\end{equation}
\end{proposition}

\begin{proof}
By Lemma~\ref{lem:J-weighted},
\begin{equation}
M
:=\sup_{|z-E_*|\le h/2}\sup_i
\sum_{k\ne i}\|J_{ik}(z)\|^s a^{s\theta(|i|-|k|)}<\infty.
\label{eq:MBtheta}
\end{equation}
Since $b_i\to0$, choose $k_0$ and
\[
\mathcal C_0:=\{i:|i|\le k_0\}
\]
such that
\begin{equation}
\sup_{|z-E_*|\le h/2}\sup_{i\notin\mathcal C_0}
 b_i\sum_{k\ne i}\|J_{ik}(z)\|^s a^{s\theta(|i|-|k|)}
\le\frac12.
\label{eq:contraction-center}
\end{equation}
Set
\begin{equation}
f_i^\Lambda:=a^{s\theta|i|}\mathbb E\|K_{i0}^\Lambda\|^s.
\label{eq:fiLambda}
\end{equation}
From \eqref{eq:uniform-K} and $\|A\|\le\|A\|_{\HS}$,
\begin{equation}
f_i^\Lambda\le A_Ba^{s\theta k_0}=:A_0,
\qquad i\in\mathcal C_0.
\label{eq:core-bound}
\end{equation}
For $i\notin\mathcal C_0$, by Lemma~\ref{lem:block-recursion},
\begin{equation}
f_i^\Lambda
\le
\sum_{k\in\Lambda\setminus\{i\}}
 b_i\|J_{ik}(z)\|^s a^{s\theta(|i|-|k|)}
 f_k^{\Lambda\setminus\{i\}}.
\label{eq:weighted-deletion}
\end{equation}
We argue by induction on $m=|\Lambda\setminus\mathcal C_0|$. For $m=0$,
\eqref{eq:core-bound} applies. Assume
\[
\sup_k f_k^{\Lambda'}\le A_0
\qquad
\text{whenever }|\Lambda'\setminus\mathcal C_0|\le m-1.
\]
If $|\Lambda\setminus\mathcal C_0|=m$ and $i\in\Lambda\setminus\mathcal C_0$, then
\[
| (\Lambda\setminus\{i\})\setminus\mathcal C_0|=m-1,
\]
so \eqref{eq:weighted-deletion} and \eqref{eq:contraction-center} give
\[
f_i^\Lambda\le\frac12A_0.
\]
Together with \eqref{eq:core-bound},
\begin{equation}
\sup_{\Lambda,i}f_i^\Lambda\le A_0.
\label{eq:uniform-f}
\end{equation}
This proves \eqref{eq:channel-decay}.

Let $\Lambda_n\uparrow\bbZ$ and set
\[
P_n=P_{B\cup W_{\Lambda_n}}.
\]
Define
\[
\widetilde H_n:=P_nHP_n\quad\text{on }\ell^2(\bbZ).
\]
Since $H$ is bounded and $P_n\to I$ strongly, for every
$u\in\ell^2(\bbZ)$,
\[
\begin{aligned}
\|\widetilde H_nu-Hu\|
&\le \|H\|\,\|(P_n-I)u\|+\|(P_n-I)Hu\|\longrightarrow0.
\end{aligned}
\]
Thus $\widetilde H_n\to H$ strongly. By the resolvent identity,
\[
(\widetilde H_n-z)^{-1}-(H-z)^{-1}
=(\widetilde H_n-z)^{-1}(H-\widetilde H_n)(H-z)^{-1}.
\]
Since $\widetilde H_n$ is self-adjoint,
\[
\sup_n\|(\widetilde H_n-z)^{-1}\|\le|\im z|^{-1},
\qquad z\in\bbC\setminus\bbR.
\]
Hence, for every $u\in\ell^2(\bbZ)$,
\[
\|(\widetilde H_n-z)^{-1}u-(H-z)^{-1}u\|
\le |\im z|^{-1}\|(H-\widetilde H_n)(H-z)^{-1}u\|\longrightarrow0.
\]
Therefore
\[
(\widetilde H_n-z)^{-1}\longrightarrow(H-z)^{-1}
\qquad\text{strongly},\qquad z\in\bbC\setminus\bbR.
\]
For sufficiently large $n$,
\begin{align}
&\left\|
K_{i0}^{\Lambda_n}(z)
-F_iP_{W_i}(H-z)^{-1}P_{W_0}F_0^*
\right\|\notag\\*
&\quad=
\left\|
F_iP_{W_i}
\bigl[(\widetilde H_n-z)^{-1}-(H-z)^{-1}\bigr]
P_{W_0}F_0^*
\right\|\notag\\*
&\quad\le
\|F_i\|\,\|F_0\|
\left(
\sum_{x\in W_0}
\left\|
\bigl[(\widetilde H_n-z)^{-1}-(H-z)^{-1}\bigr]\delta_x
\right\|^2
\right)^{1/2}
\longrightarrow0.
\label{eq:K-finite-to-infinite}
\end{align}
Consequently,
\[
\left\|
F_iP_{W_i}(H-z)^{-1}P_{W_0}F_0^*
\right\|^s
=\lim_{n\to\infty}\|K_{i0}^{\Lambda_n}(z)\|^s.
\]
Taking expectation and using Fatou's lemma together with
\eqref{eq:channel-decay},
\[
\mathbb E\left\|F_iP_{W_i}(H-z)^{-1}P_{W_0}F_0^*\right\|^s
=\mathbb E\!\left[\lim_{n\to\infty}\|K_{i0}^{\Lambda_n}(z)\|^s\right]
\le\liminf_{n\to\infty}\mathbb E\|K_{i0}^{\Lambda_n}(z)\|^s
\le Ca^{-s\theta|i|},
\]
which is \eqref{eq:channel-decay-infinite}.
\end{proof}

\section{Spectral localization}
\label{sec:terminal}
\label{sec:simon-wolff}
\label{sec:parameters}

Fix $x\in\bbZ$ and $0\le\tau<\alpha$. Enlarging $N$ in Lemma~\ref{lem:black-union-gap} if necessary,
assume
\begin{equation}
x\in W_0,
\qquad
\dist(E_*,\sigma(H_B))\ge6g.
\label{eq:x-in-W0}
\end{equation}
Let $I$ be as in \eqref{eq:local-z}.
For $z\in\bbC\setminus\bbR$, set
\[
G(n,m;z):=\langle\delta_n,(H-z)^{-1}\delta_m\rangle.
\]
Let $\mu_x$ denote the spectral measure of $\delta_x$:
\[
\mu_x(S):=\langle\delta_x,P_S(H)\delta_x\rangle,
\qquad S\subset\bbR\ \text{Borel}.
\]

\begin{proposition}
\label{prop:green-square}
For $\mathbb P$-a.e. $\omega$ and Lebesgue-a.e. $E\in I$,
\begin{equation}
\lim_{\eta\downarrow0}
\sum_{n\in\bbZ}|G(n,x;E+i\eta)|^2
=
\int_{\bbR}\frac{\mu_x(dt)}{(t-E)^2}
<\infty.
\label{eq:green-square}
\end{equation}
Moreover,
\begin{equation}
(H-E-i\eta)^{-1}\delta_x
\xrightarrow[\eta\downarrow0]{\ell^2}
(H-E-i0)^{-1}\delta_x\in\ell^2_\tau(\bbZ).
\label{eq:full-l2-boundary}
\end{equation}

\end{proposition}

\begin{proof}
We first work conditionally on $\cF_B$, with expectation as in
Section~\ref{sec:schur-factorization}.
Choose
\begin{equation}
\tau+\frac12<\theta_0<\theta_1<\theta_2<\beta.
\label{eq:theta-chain}
\end{equation}

To apply Proposition~\ref{prop:channel-decay}, define
\[
\widetilde{\cK}_0:=\cK_0\oplus\ell^2(W_0),
\qquad
\widetilde F_0:\ell^2(W_0)\to\widetilde{\cK}_0,
\qquad \widetilde F_0u:=(F_0u,u).
\]
Then
\[
\widetilde F_0^*(v,w)=F_0^*v+w,
\qquad
\Phi(\widetilde F_0)^{p_0}
=\sum_{n\in W_0}\bigl(\|F_0\delta_n\|^2+1\bigr)^{p_0}
\le\Phi(F_0)^{p_0}+|W_0|<\infty.
\]
Set $\widetilde J(z)=J(z)\oplus0$ on
$\widetilde{\cK}=\cK\oplus\ell^2(W_0)$, and let $\widetilde F_i=F_i$ for $i\ne0$.
Then \eqref{eq:V-FJF} remains valid:
\[
V(z)=\widetilde F^*\widetilde J(z)\widetilde F.
\]
Since only $F_0$ changes and $\Phi(\widetilde F_0)<\infty$, we still have
\[
\sup_i\Phi(\widetilde F_i)<\infty,
\qquad \Phi(\widetilde F_i)\to0\quad(|i|\to\infty).
\]
Moreover, $\|\widetilde J_{ij}(z)\|=\|J_{ij}(z)\|$, so
Lemma~\ref{lem:J-weighted} remains valid.
Thus Lemma~\ref{lem:apriori} and Proposition~\ref{prop:channel-decay}
remain valid for the enlarged family, with possibly different constants.
Henceforth we omit the tildes and use $\cK_0,F_0,F,\cK,J(z)$ for the
enlarged spaces and operators. In particular, let
\begin{equation}
e_x:=(0,\delta_x)\in\cK_0\subset\cK,
\qquad
F^*e_x=F_0^*e_x=\delta_x.
\label{eq:ex-source}
\end{equation}
For $z=E+i\eta$ as in \eqref{eq:local-z}, set
\[
u(z):=(H-z)^{-1}\delta_x,
\qquad
u_W(z):=P_Wu(z),
\qquad
u_B(z):=P_Bu(z).
\]
Define
\begin{equation}
q_i(z):=F_iP_{W_i}u(z),
\qquad
q(z):=(q_i(z))_{i\in\bbZ}=Fu_W(z).
\label{eq:q-definition}
\end{equation}
By \eqref{eq:schur-inverse}, \eqref{eq:schur}, and \eqref{eq:V-FJF}, together
with $q(z)=Fu_W(z)$ and $F^*e_x=\delta_x$ from
\eqref{eq:ex-source}--\eqref{eq:q-definition}, we have
\[
D(z)u_W(z)=\delta_x-F^*J(z)Fu_W(z)
=F^*\bigl(e_x-J(z)q(z)\bigr).
\]
Since $\im D_i(z)\le-\eta I_{W_i}$ for every $i$, we have
$\|D(z)^{-1}\|\le\eta^{-1}$. Hence
\begin{equation}
u_W(z)
=
\mathcal T(z)\bigl(e_x-J(z)q(z)\bigr),
\qquad
\mathcal T(z):=D(z)^{-1}F^*.
\label{eq:uW}
\end{equation}
We first obtain a weighted subsequential limit for $q(z)$ and a weighted boundary
limit for $\mathcal T(z)$. Equation~\eqref{eq:uW} then yields a weighted
subsequential limit for $u_W(z)$, and the $B$-row of the resolvent equation
recovers $u_B(z)$. Finally, the spectral theorem gives the conclusions of the proposition.

\smallskip
\noindent\emph{Step 1. Weighted subsequential convergence of $q(z)$.}
By \eqref{eq:ex-source},
\[
\|q_i(z)\|\le\|F_iP_{W_i}(H-z)^{-1}P_{W_0}F_0^*\|.
\]
Hence, by \eqref{eq:channel-decay-infinite} with $\theta=\theta_2$ as chosen in
\eqref{eq:theta-chain},
\begin{equation}
\mathbb E\|q_i(E+i\eta)\|^s
\le C_Ba^{-s\theta_2|i|}.
\label{eq:q-block-moment}
\end{equation}
Since $\theta_1<\theta_2$, \eqref{eq:q-block-moment} yields
\begin{align}
\sup_{\substack{E\in I\\0<\eta<h/4}}
\mathbb E
\sum_i a^{s\theta_1|i|}\|q_i(E+i\eta)\|^s
&\le
C_B\sum_i a^{-s(\theta_2-\theta_1)|i|}
<\infty.
\label{eq:q-s-moment}
\end{align}

Fix any sequence $\varepsilon_n\downarrow0$ and set
\[
X_n(\omega,E):=
\sum_i a^{s\theta_1|i|}\|q_i(E+i\varepsilon_n)\|^s.
\]
By Tonelli's theorem, Fatou's lemma, and \eqref{eq:q-s-moment},
\begin{equation}
\int_I\mathbb E\!\left[\liminf_{n\to\infty}X_n(\omega,E)\right]dE
\le
\liminf_{n\to\infty}\int_I\mathbb E X_n(\omega,E)\,dE
<\infty.
\label{eq:fatou-channel}
\end{equation}
\begin{equation}
\begin{gathered}
\text{For }(\mathbb P_B\otimes dE)\text{-a.e. }(\omega,E),\quad
\exists\,n_k=n_k(\omega,E)\uparrow\infty\text{ such that}\\
X_{n_k}(\omega,E)\longrightarrow\liminf_{n\to\infty}X_n(\omega,E)<\infty,
\qquad\sup_k X_{n_k}(\omega,E)<\infty.
\end{gathered}
\label{eq:channel-subsequence-bound}
\end{equation}
Since $0<s<2/3$,
\begin{equation}
\begin{aligned}
\|q(E+i\varepsilon_{n_k})\|_{X_{\theta_1}}
&=\left(\sum_i a^{2\theta_1|i|}\|q_i(E+i\varepsilon_{n_k})\|^2\right)^{1/2}\\
&\le\left(\sum_i a^{s\theta_1|i|}\|q_i(E+i\varepsilon_{n_k})\|^s\right)^{1/s}
=X_{n_k}(\omega,E)^{1/s}.
\end{aligned}
\label{eq:lsp-to-l2}
\end{equation}
Hence $\{q(E+i\varepsilon_{n_k})\}_k$ is bounded in $X_{\theta_1}$.

For each $i$, set
\[
M_1=F_iP_{W_i},
\qquad
M_2:\bbC\to\ell^2(\bbZ),\quad M_2c=c\delta_x.
\]
\[
\|M_1\|_{\HS}^2=\sum_{n\in W_i}\|F_i\delta_n\|^2<\infty,
\qquad\|M_2\|_{\HS}^2=1.
\]
By Lemma~\ref{lem:weakL1},
\begin{equation}
q_i(E+i0):=
\lim_{\eta\downarrow0}q_i(E+i\eta)
\quad\text{exists in }\cK_i
\quad\text{for Lebesgue-a.e. }E.
\label{eq:blockwise-q-boundary}
\end{equation}
Since $\bbZ$ is countable, the limits in \eqref{eq:blockwise-q-boundary}
exist for every $i\in\bbZ$ on a common set of full Lebesgue measure in $I$.
For such $E$, write $q(E):=(q_i(E+i0))_{i\in\bbZ}$. By \eqref{eq:channel-subsequence-bound} and Fatou's lemma,
\begin{align}
\|q(E)\|_{X_{\theta_1}}^2
&=\sum_i a^{2\theta_1|i|}\|q_i(E+i0)\|^2\notag\\
&\le\liminf_{k\to\infty}\sum_i a^{2\theta_1|i|}\|q_i(E+i\varepsilon_{n_k})\|^2\notag\\
&=\liminf_{k\to\infty}\|q(E+i\varepsilon_{n_k})\|_{X_{\theta_1}}^2<\infty.
\label{eq:q-limit-Xtheta1}
\end{align}
By \eqref{eq:blockwise-q-boundary}, for each fixed $R\ge1$,
\begin{align*}
&\limsup_{k\to\infty}\|q(E+i\varepsilon_{n_k})-q(E)\|_{X_{\theta_0}}^2\\*
&\quad=\limsup_{k\to\infty}
\sum_{|i|>R}a^{2\theta_0|i|}
\|q_i(E+i\varepsilon_{n_k})-q_i(E)\|^2\\
&\quad\le 2a^{-2(\theta_1-\theta_0)R}
\left(\sup_k\|q(E+i\varepsilon_{n_k})\|_{X_{\theta_1}}^2+\|q(E)\|_{X_{\theta_1}}^2\right)\\
&\quad\le Ca^{-2(\theta_1-\theta_0)R}\xrightarrow[R\to\infty]{}0.
\end{align*}
It follows that
\begin{equation}
q(E+i\varepsilon_{n_k})\longrightarrow q(E)
\qquad\text{in }X_{\theta_0}.
\label{eq:q-boundary}
\end{equation}

\smallskip
\noindent\emph{Step 2. Weighted boundary convergence of $\mathcal T(z)$.}
Let $\mathfrak Y:=\bigoplus_i\ell^2(W_i)$ and define
\begin{equation}
\begin{gathered}
Y_\tau:=\left\{u=(u_{W_i})_i\in\mathfrak Y:
\sum_i a^{2\tau|i|}\|u_{W_i}\|^2<\infty\right\},\\
\|u\|_{Y_\tau}^2:=\sum_i a^{2\tau|i|}\|u_{W_i}\|^2.
\end{gathered}
\label{eq:Ytau}
\end{equation}
We will prove that $\mathcal T(E+i\eta)$ converges in
$\mathcal B(X_{\theta_0},Y_\tau)$ as $\eta\downarrow0$ for Lebesgue-a.e. $E$.
To apply Lemma~\ref{lem:weakL1}, recall $D_i(z)$ from
\eqref{eq:D-blocks} and set
\begin{equation}
\widehat H_i:=H_{B\cup W_i},
\qquad
R_i(z):=D_i(z)^{-1}
=P_{W_i}(\widehat H_i-z)^{-1}P_{W_i}.
\label{eq:Ri}
\end{equation}
Define
\[
d_i:=(1+|i|)^{-1},
\qquad
\mathfrak H:=\bigoplus_i\ell^2(B\cup W_i),
\qquad
\mathfrak A:=\bigoplus_i\widehat H_i.
\]
Each $\widehat H_i$ is self-adjoint, and
\[
\sup_i\|\widehat H_i\|
\le 2\sum_{n\ge1}n^{-r}+\lambda.
\]
Consequently $\mathfrak A$ is a bounded self-adjoint operator on $\mathfrak H$.
Define
\begin{equation}
M_1:=\bigoplus_i a^{-(\theta_0-\tau)|i|}d_i^{-1}P_{W_i}:
\mathfrak H\to\mathfrak Y,
\qquad
M_2:=\bigoplus_i d_iP_{W_i}F_i^*:\cK\to\mathfrak H.
\label{eq:M1M2}
\end{equation}
In $M_2$, vectors $F_i^*v$, $v\in\cK_i$, are extended by zero from $W_i$ to $B\cup W_i$.
By \eqref{eq:phi-hs}, \eqref{eq:white-size}, \eqref{eq:Fi-decay}, and
\eqref{eq:theta-chain},
\begin{align}
\|M_2\|_{\HS}^2
&=\sum_i d_i^2\|F_i\|_{\HS}^2
\le\sum_i d_i^2\Phi(F_i)
\le C_B\sum_i(1+|i|)^{-2-\alpha}<\infty,
\label{eq:M2-HS}\\
\|M_1\|_{\HS}^2
&=\sum_i a^{-2(\theta_0-\tau)|i|}d_i^{-2}N_i
\le C_B\sum_i(1+|i|)^2a^{-[2(\theta_0-\tau)-1]|i|}<\infty.
\label{eq:M1-HS}
\end{align}
By Lemma~\ref{lem:weakL1} and \eqref{eq:M2-HS}--\eqref{eq:M1-HS}, for Lebesgue-a.e. $E$,
\begin{equation}
Z(E+i\eta)
:=M_1(\mathfrak A-E-i\eta)^{-1}M_2
=
\bigoplus_i a^{-(\theta_0-\tau)|i|}R_i(E+i\eta)F_i^*
\xrightarrow[\eta\downarrow0]{\HS} Z(E).
\label{eq:terminal-HS}
\end{equation}
Define the isometric isomorphisms
\[
\begin{aligned}
U_X:X_{\theta_0}\to\cK,&\qquad (U_Xq)_i=a^{\theta_0|i|}q_i,\\
U_Y:Y_\tau\to\mathfrak Y,&\qquad (U_Yu)_i=a^{\tau|i|}u_{W_i}.
\end{aligned}
\]
Then
\begin{equation}
U_Y\mathcal T(z)U_X^{-1}=Z(z).
\label{eq:terminal-conjugation}
\end{equation}
For $E$ satisfying \eqref{eq:terminal-HS}, define
\[
\mathcal T(E):=U_Y^{-1}Z(E)U_X.
\]
By \eqref{eq:terminal-HS} and \eqref{eq:terminal-conjugation},
\begin{align}
&\|\mathcal T(E+i\eta)-\mathcal T(E)\|_{\mathcal B(X_{\theta_0},Y_\tau)}\notag\\*
&\qquad=\|Z(E+i\eta)-Z(E)\|_{\mathcal B(\cK,\mathfrak Y)}\notag\\*
&\qquad\le
\|Z(E+i\eta)-Z(E)\|_{\HS}
\xrightarrow[\eta\downarrow0]{}0.
\label{eq:D-inverse-terminal}
\end{align}
\smallskip
\noindent\emph{Step 3. Weighted subsequential convergence of $u(z)$.}
In what follows, fix $(\omega,E)$ in the full
$\mathbb P_B\otimes\Leb$-measure set on which
\eqref{eq:channel-subsequence-bound}, \eqref{eq:blockwise-q-boundary},
and \eqref{eq:terminal-HS} hold.

By \eqref{eq:q-boundary} and Lemma~\ref{lem:J-weighted},
\[
J(E+i\varepsilon_{n_k})q(E+i\varepsilon_{n_k})\longrightarrow J(E)q(E)
\qquad\text{in }X_{\theta_0}.
\]
Together with \eqref{eq:D-inverse-terminal} and \eqref{eq:uW}, this gives
\begin{equation}
\begin{aligned}
u_W(E+i\varepsilon_{n_k})
&=\mathcal T(E+i\varepsilon_{n_k})
\bigl(e_x-J(E+i\varepsilon_{n_k})q(E+i\varepsilon_{n_k})\bigr)\\
&\longrightarrow
\mathcal T(E)\bigl(e_x-J(E)q(E)\bigr)=:u_W(E)
\qquad\text{in }Y_\tau.
\end{aligned}
\label{eq:uW-limit}
\end{equation}
By \eqref{eq:white-size},
\begin{equation}
\|v\|_{\ell^2_\tau(W)}^2
=\sum_i\sum_{n\in W_i}\langle n\rangle^{2\tau}|v(n)|^2
\le C_B\sum_i a^{2\tau|i|}\|v_{W_i}\|^2
=C_B\|v\|_{Y_\tau}^2.
\label{eq:Ytau-physical}
\end{equation}
By \eqref{eq:uW-limit} and \eqref{eq:Ytau-physical},
\begin{equation}
u_W(E+i\varepsilon_{n_k})\longrightarrow u_W(E)
\quad\text{in }\ell^2_\tau(W).
\label{eq:uW-weighted-limit}
\end{equation}
The $B$-row of $(H-z)u=\delta_x$ gives
\begin{equation}
u_B(z)=-R_B(z)T_{BW}u_W(z).
\label{eq:uB}
\end{equation}
Since $\tau<\alpha<r-1$, Corollary~\ref{cor:weighted-J} gives
\[
\|T_{BW}\|_{\mathcal B(\ell^2_\tau(W),\ell^2_\tau(B))}
\le C\|T_r\|_{J_r}\le C.
\]
Moreover, by Lemma~\ref{lem:black-union-gap}, Proposition~\ref{prop:uniform-inverse},
and Corollary~\ref{cor:weighted-J},
\begin{equation}
\sup_{|z-E_*|\le h/2}\|R_B(z)\|_{\mathcal B(\ell^2_\tau(B))}\le C_B.
\label{eq:weighted-B-bounds}
\end{equation}
For $|z-E_*|,|w-E_*|\le h/2$, the resolvent identity gives
\begin{equation}
\|R_B(z)-R_B(w)\|_{\mathcal B(\ell^2_\tau(B))}
=|z-w|\|R_B(z)R_B(w)\|_{\mathcal B(\ell^2_\tau(B))}
\le C_B|z-w|.
\label{eq:RB-weighted-continuity}
\end{equation}
Define
\[
u_B(E):=-R_B(E)T_{BW}u_W(E).
\]
Then, by \eqref{eq:uW-weighted-limit}--\eqref{eq:RB-weighted-continuity},
\begin{equation}
\begin{aligned}
&\|u_B(E+i\varepsilon_{n_k})-u_B(E)\|_{\ell^2_\tau(B)}\\
&\qquad\le C_B\|u_W(E+i\varepsilon_{n_k})-u_W(E)\|_{\ell^2_\tau(W)}
+C_B\varepsilon_{n_k}\|u_W(E)\|_{\ell^2_\tau(W)}\longrightarrow0.
\end{aligned}
\label{eq:uB-weighted-limit}
\end{equation}
Combining \eqref{eq:uW-weighted-limit} and \eqref{eq:uB-weighted-limit}, and setting
$u(E):=u_W(E)+u_B(E)$, we obtain
\begin{equation}
u(E+i\varepsilon_{n_k})\longrightarrow u(E)
\qquad\text{in }\ell^2_\tau(\bbZ).
\label{eq:weighted-subsequence-limit}
\end{equation}

\smallskip
\noindent\emph{Step 4. Completion of the proof.}
The spectral theorem gives
\begin{equation}
\|(H-E-i\eta)^{-1}\delta_x\|^2
=
\int_{\bbR}\frac{\mu_x(dt)}{(t-E)^2+\eta^2}.
\label{eq:spectral-resolvent-norm}
\end{equation}
By \eqref{eq:weighted-subsequence-limit}, \eqref{eq:spectral-resolvent-norm}, and the monotone convergence theorem,
\begin{equation}
\begin{aligned}
\int_{\bbR}\frac{\mu_x(dt)}{(t-E)^2}
&=\lim_{\eta\downarrow0}\int_{\bbR}\frac{\mu_x(dt)}{(t-E)^2+\eta^2}\\
&=\lim_{k\to\infty}\|(H-E-i\varepsilon_{n_k})^{-1}\delta_x\|_{\ell^2}^2
=\|u(E)\|_{\ell^2}^2<\infty.
\end{aligned}
\label{eq:B-finite}
\end{equation}
Furthermore,
\[
\left|\frac1{t-E-i\eta}-\frac1{t-E}\right|^2
\le\frac{4}{(t-E)^2}.
\]
By the dominated convergence theorem in $L^2(\mu_x)$ and \eqref{eq:weighted-subsequence-limit},
\[
\begin{aligned}
(H-E-i0)^{-1}\delta_x
&=\lim_{\eta\downarrow0}(H-E-i\eta)^{-1}\delta_x\\
&=\lim_{k\to\infty}(H-E-i\varepsilon_{n_k})^{-1}\delta_x
=u(E)\in\ell^2_\tau(\bbZ),
\end{aligned}
\]
where both limits are taken in $\ell^2$. This proves \eqref{eq:full-l2-boundary}.

For each fixed $E_*$, $x\in\bbZ$, and $0\le\tau<\alpha$, let
\[
\mathcal N:=\{(\omega,E):E\in I,\ \eqref{eq:green-square}
\text{ or }\eqref{eq:full-l2-boundary}\text{ fails}\}.
\]
The conditional argument above shows that
\begin{equation}
(\mathbb P_B\otimes\Leb)(\mathcal N)=0\quad\text{almost surely}.
\label{eq:conditional-failure-zero}
\end{equation}
Taking unconditional expectations in \eqref{eq:conditional-failure-zero} gives
\[
(\mathbb P_0\otimes\Leb)(\mathcal N)=0,
\]
since $\mathcal N$ depends only on the original disorder variables and their marginal law is $\mathbb P_0$.
This proves both assertions.
\end{proof}

Henceforth, write $\mathbb P=\mathbb P_0$ and set
\begin{equation}
\Sigma_\lambda
:=
\left[-2\sum_{k\ge1}k^{-r}-\lambda,
2\sum_{k\ge1}k^{-r}+\lambda\right],
\label{eq:det-spectrum-window}
\end{equation}
so that $\sigma(H_{\lambda,\omega})\subseteq\Sigma_\lambda$.
For each $E_*\in\Sigma_\lambda$, Lemma~\ref{lem:taylor} provides a deterministic
$h(E_*)>0$ and
\[
I_{E_*}=(E_*-h(E_*)/4,E_*+h(E_*)/4).
\]
Compactness gives
\begin{equation}
\Sigma_\lambda\subset\bigcup_{m=1}^M I_{E_m}.
\label{eq:finite-energy-cover}
\end{equation}
For $E\notin\Sigma_\lambda$, \eqref{eq:det-spectrum-window} gives $E\in\rho(H)$,
while Proposition~\ref{prop:uniform-inverse} and Corollary~\ref{cor:weighted-J} yield
\[
(H-E)^{-1}\delta_x\in\ell^2_\tau(\bbZ),\qquad 0\le\tau<r-1.
\]
Therefore, after a countable intersection over
$x\in\bbZ$ and over any sequence $0\le\tau_k<\alpha$ with $\tau_k\uparrow\alpha$, there is a
probability-one event on which, for every $x$ and $k$,
\begin{equation}
\int\frac{\mu_x(dt)}{(t-E)^2}<\infty,
\qquad
(H-E-i0)^{-1}\delta_x\in\ell^2_{\tau_k}
\label{eq:global-good-energy}
\end{equation}
for Lebesgue-a.e. $E\in\bbR$.

\begin{lemma}\label{lem:rank-one}
Let $\cH$ be a separable Hilbert space, and let $\cH_\tau\subseteq\cH$ be a linear subspace equipped with a norm such that
\[
\|u\|_{\cH}\le C_\tau\|u\|_{\cH_\tau},\qquad u\in\cH_\tau,
\]
for some $C_\tau<\infty$. Let $A=A^*\in\mathcal B(\cH)$ and $e\in\cH$ with $\|e\|_{\cH}=1$.
Set
\[
A_t=A+t|e\ra\la e|
\]
and
\[
\mu_{t,e}(S):=\la e,P_S(A_t)e\ra,\qquad S\subset\bbR\ \text{Borel}.
\]
Assume that for a Borel set $G\subset\bbR$ with $\Leb(G^c)=0$,
\begin{equation}
\begin{gathered}
B(E):=\int\frac{\mu_{0,e}(du)}{(u-E)^2}<\infty,
\\
u_E:=\lim_{\eta\downarrow0}(A-E-i\eta)^{-1}e
=(A-E-i0)^{-1}e\in\cH_\tau,
\qquad E\in G,
\end{gathered}
\label{eq:rank-one-hyp}
\end{equation}
where the limit is taken in $\cH$. Then for Lebesgue-a.e. $t$,
$\mu_{t,e}$ is pure point and
\begin{equation}
P_{\{E\}}(A_t)e\in\cH_\tau\qquad\text{for every }E\in\bbR.
\label{eq:rank-one-weighted-conclusion}
\end{equation}
\end{lemma}

\begin{proof}
For $z\in\bbC\setminus\bbR$, set
\[
m(z):=\la e,(A-z)^{-1}e\ra,
\qquad
m_t(z):=\la e,(A_t-z)^{-1}e\ra.
\]
The rank-one resolvent identity \cite[Theorem~1.1.1]{AlbeverioKurasov2000} gives
\begin{equation}
m_t(z)=\frac{m(z)}{1+tm(z)},
\qquad
(A_t-z)^{-1}e
=\frac{(A-z)^{-1}e}{1+tm(z)}.
\label{eq:krein}
\end{equation}
For $E\in G$,
\[
m(E):=\int\frac{\mu_{0,e}(du)}{u-E}\in\bbR
\]
and
\begin{align}
\frac{m(E+i\eta)-m(E)}{i\eta}
&=
\int\frac{\mu_{0,e}(du)}{(u-E)(u-E-i\eta)}
\xrightarrow[\eta\downarrow0]{} B(E),
\label{eq:m-derivative}\\
m(E+i\eta)&=m(E)+i\eta B(E)+o(\eta).
\label{eq:m-expansion}
\end{align}
By \eqref{eq:rank-one-hyp},
\begin{equation}
(A-E)u_E=e,
\qquad
\la e,u_E\ra=m(E).
\label{eq:uE-equation}
\end{equation}
For $t\ne0$, define
\[
\mathcal R_t:=\{E\in G:1+tm(E)=0\}.
\]
If $E\in\mathcal R_t$, then by \eqref{eq:uE-equation}
\[
(A_t-E)u_E=(1+tm(E))e=0.
\]
Thus $m(E)=-1/t$, and \eqref{eq:krein}--\eqref{eq:m-expansion} give
\[
\lim_{\eta\downarrow0}\eta m_t(E+i\eta)
=\lim_{\eta\downarrow0}\frac{\eta m(E+i\eta)}{1+tm(E+i\eta)}
=\frac{-1/t}{itB(E)}=\frac{i}{t^2B(E)}.
\]
By the dominated convergence theorem,
\begin{equation}
\begin{aligned}
\mu_{t,e}(\{E\})
&=\lim_{\eta\downarrow0}\int_{\bbR}
\frac{\eta^2}{(u-E)^2+\eta^2}\,\mu_{t,e}(du)\\
&=\lim_{\eta\downarrow0}\eta\,\im m_t(E+i\eta)
=\frac1{t^2B(E)}>0.
\end{aligned}
\label{eq:atom-mass}
\end{equation}
If $E\in G\setminus\mathcal R_t$, then $1+tm(E)\ne0$. Since
\[
\im m_t(E+i\eta)=\frac{\im m(E+i\eta)}{|1+tm(E+i\eta)|^2},
\]
it follows from \eqref{eq:m-expansion} that
\begin{equation}
\lim_{\eta\downarrow0}\im m_t(E+i\eta)=0.
\label{eq:nonresonant-zero}
\end{equation}
Write the Lebesgue decomposition of $\mu_{t,e}$ as
\[
\mu_{t,e}=\mu_{t,e}^{\rm ac}+\mu_{t,e}^{\rm s},
\qquad
\mu_{t,e}^{\rm ac}\ll\Leb,
\quad
\mu_{t,e}^{\rm s}\perp\Leb.
\]
By \cite[Proposition~3.2]{Hislop2008Lectures}, the density of $\mu_{t,e}^{\rm ac}$ is given Lebesgue-almost everywhere by $\pi^{-1}\im m_t(E+i0)$, and $\mu_{t,e}^{\rm s}$ is concentrated on
\[
\left\{E:\lim_{\eta\downarrow0}\im m_t(E+i\eta)=+\infty\right\}.
\]
In particular,
\begin{equation}
\mu_{t,e}^{\rm s}\left(\bbR\setminus
\left\{E:\lim_{\eta\downarrow0}\im m_t(E+i\eta)=+\infty\right\}\right)=0.
\label{eq:borel-boundary-decomposition}
\end{equation}
Thus,
\[
\mu_{t,e}^{\rm ac}(G\setminus\mathcal R_t)=0,
\qquad
\mu_{t,e}^{\rm s}(G\setminus\mathcal R_t)=0.
\]
Hence $\mu_{t,e}(G\setminus\mathcal R_t)=0$. On the other hand,
\eqref{eq:atom-mass} shows that every point of $\mathcal R_t$ is an atom.
Therefore
\begin{equation}
\mu_{t,e}|_G
=
\sum_{E\in\mathcal R_t}\mu_{t,e}(\{E\})\,\delta_E.
\label{eq:mu-on-G}
\end{equation}
By the rank-one spectral averaging formula \cite[Theorem~I.8]{Simon1995RankOne}, for every Borel set $B\subset\bbR$,
\begin{equation}
\int_{\bbR}\mu_{t,e}(B)\,dt=\Leb(B).
\label{eq:rank-one-averaging}
\end{equation}
Hence
\[
\int_{\bbR}\mu_{t,e}(G^c)\,dt=0,
\qquad
\mu_{t,e}(G^c)=0
\quad\text{for a.e. }t.
\]
Together with \eqref{eq:mu-on-G}, this shows that $\mu_{t,e}$ is a pure point measure for Lebesgue-almost every $t\in\bbR$.

Finally, fix $t\ne0$ such that $\mu_{t,e}(G^c)=0$.
For $E\in\mathcal R_t$, \eqref{eq:m-expansion} gives
\[
\frac{-i\eta}{1+tm(E+i\eta)}
=\frac{-i\eta}{it\eta B(E)+o(\eta)}
\xrightarrow[\eta\downarrow0]{}-\frac1{tB(E)}.
\]
Thus, by the spectral theorem and \eqref{eq:krein},
\begin{align*}
P_{\{E\}}(A_t)e
&=\lim_{\eta\downarrow0}(-i\eta)(A_t-E-i\eta)^{-1}e\\
&=\lim_{\eta\downarrow0}
\frac{-i\eta}{1+tm(E+i\eta)}(A-E-i\eta)^{-1}e\\
&=-\frac1{tB(E)}u_E\in\cH_\tau,
\end{align*}
where both vector limits are taken in $\cH$.
For $E\notin\mathcal R_t$, $P_{\{E\}}(A_t)e=0$, since $\mu_{t,e}$ is concentrated on $\mathcal R_t$.
\end{proof}

\begin{proof}[Proof of Theorem~\ref{thm:main}]
Fix a sequence $0\le\tau_k<\alpha$ with $\tau_k\uparrow\alpha$,
and fix $x\in\bbZ$ and $k\ge1$.
Condition on the random variables $\omega_n$, $n\ne x$, and write
\begin{equation}
H_t:=H_0+t|\delta_x\ra\la\delta_x|,
\qquad H_0:=H\big|_{\lambda\omega_x=0}.
\label{eq:Ht-family}
\end{equation}
Let $\mu_{t,x}$ denote the spectral measure of $\delta_x$ for $H_t$.
Fubini's theorem and \eqref{eq:global-good-energy} imply that, for almost every
$\omega_{\ne x}$, $H_t$ satisfies \eqref{eq:global-good-energy} for Lebesgue-a.e.
$t\in[-\lambda,\lambda]$.

Fix such an $\omega_{\ne x}$ and choose $t_*\in(-\lambda,\lambda)$ for which
\eqref{eq:global-good-energy} holds. There exists a Borel set
$G_*\subset\bbR$ with $\Leb(G_*^c)=0$ such that, for every $E\in G_*$,
\[
\begin{gathered}
\int_{\bbR}\frac{\mu_{t_*,x}(du)}{(u-E)^2}<\infty,
\\
(H_{t_*}-E-i\eta)^{-1}\delta_x
\xrightarrow[\eta\downarrow0]{\ell^2(\bbZ)}
(H_{t_*}-E-i0)^{-1}\delta_x\in\ell^2_{\tau_k}(\bbZ).
\end{gathered}
\]
Thus Lemma~\ref{lem:rank-one} applies with
\[
A=H_{t_*},\qquad e=\delta_x,\qquad
\cH=\ell^2(\bbZ),\qquad\cH_\tau=\ell^2_{\tau_k}.
\]
Since
\[
H_t=H_{t_*}+(t-t_*)|\delta_x\ra\la\delta_x|,
\]
the lemma implies that, for Lebesgue-a.e. $t\in[-\lambda,\lambda]$,
$\mu_{t,x}$ is pure point and
\[
P_{\{E\}}(H_t)\delta_x\in\ell^2_{\tau_k}\qquad\text{for every }E\in\bbR.
\]
Define
\[
\mathcal A_{x,k}
:=\left\{(\omega,E):
\mu_x(\{E\})>0,\quad
P_{\{E\}}(H)\delta_x\in\ell^2_{\tau_k}
\right\}.
\]
This set is measurable by the joint measurability of the point spectral projections and the definition of the weighted norm.
The event in \eqref{eq:coordinate-pure-point} is precisely
\[
\left\{\omega:
\int_{\bbR}\one_{\mathcal A_{x,k}}(\omega,E)\,\mu_x(dE)=1
\right\},
\]
which is measurable by the measurability of the spectral measures.
Since $t=\lambda\omega_x$ has an absolutely continuous distribution, Fubini's theorem yields, almost surely,
\begin{equation}
\mu_x\text{ is pure point},\qquad
P_{\{E\}}(H)\delta_x\in\ell^2_{\tau_k}\quad\text{for every }E\in\bbR.
\label{eq:coordinate-pure-point}
\end{equation}

Take the countable intersection of \eqref{eq:coordinate-pure-point} over
$x\in\bbZ$ and $k\ge1$. Then, almost surely,
\begin{equation}
P_{\rm c}(H)\delta_x=0\quad\forall x
\quad\Longrightarrow\quad
\cH_{\rm c}(H)=\{0\},
\label{eq:no-continuous-subspace}
\end{equation}
and
\begin{equation}
P_{\{E\}}(H)\delta_x
\in\bigcap_{k\ge1}\ell^2_{\tau_k}
\qquad\forall x,E.
\label{eq:projected-weighted}
\end{equation}
For every eigenvalue $E$,
\begin{equation}
\Ran P_{\{E\}}(H)
=
\overline{\Span\{P_{\{E\}}(H)\delta_x:x\in\bbZ\}}.
\label{eq:eigenspace-span}
\end{equation}
\begin{samepage}
Applying Gram--Schmidt to $\{P_{\{E\}}(H)\delta_x:x\in\bbZ\}$ in each eigenspace yields an orthonormal eigenbasis $\{\psi_j\}$. By \eqref{eq:projected-weighted}, each resulting vector, being a finite linear combination, satisfies
\begin{equation}
\psi_j\in\bigcap_{k\ge1}\ell^2_{\tau_k}
=\bigcap_{0\le\tau<\alpha}\ell^2_\tau,
\qquad
\alpha=r-\frac32.
\label{eq:final-weighted-basis}
\end{equation}
This proves Theorem~\ref{thm:main}.
\end{samepage}
\end{proof}

\appendix
\section{Proofs of the spectral-averaging estimates}
\label{app:averaging-proofs}

\begin{proof}[Proof of Lemma~\ref{lem:weakL1}]
The formulation above differs from \cite[Lemma~3.1]{AizenmanEtAl2006} only in
allowing the auxiliary Hilbert spaces $\cK_1$ and $\cK_2$ to be different. Since
$\cH$ is separable and $M_1,M_2^*$ are bounded, both
$\overline{\Ran M_1}\subset\cK_1$ and
$\overline{\Ran M_2^*}\subset\cK_2$ are separable. We may replace
$\cK_1$ and $\cK_2$ by these closed subspaces. Set
$\cK=\cK_1\oplus\cK_2$, equipped with the usual Hilbert direct-sum structure, and define
\[
\widetilde M_1:\cH\to\cK,
\qquad
\widetilde M_1u=(M_1u,0),
\]
and
\[
\widetilde M_2:\cK\to\cH,
\qquad
\widetilde M_2(x_1,x_2)=M_2x_2.
\]
Then, as an operator on $\cK$,
\[
\widetilde M_1(A-t+i\eps)^{-1}\widetilde M_2
=
\begin{pmatrix}
0 & M_1(A-t+i\eps)^{-1}M_2\\
0 & 0
\end{pmatrix}.
\]
Moreover,
$\|\widetilde M_1\|_{\HS}=\|M_1\|_{\HS}$ and
$\|\widetilde M_2\|_{\HS}=\|M_2\|_{\HS}$. Let $\{e_j\}$ and
$\{f_k\}$ be orthonormal bases of $\cK_1$ and $\cK_2$, respectively. For
$B\in\HS(\cK_2,\cK_1)$, the natural orthonormal basis $\{(e_j,0)\}_j\cup\{(0,f_k)\}_k$ of $\cK$ gives
\[
\left\|\begin{pmatrix}0&B\\0&0\end{pmatrix}\right\|_{\HS}^2
=\sum_k\|Bf_k\|_{\cK_1}^2=\|B\|_{\HS}^2.
\]
Applying \cite[Lemma~3.1]{AizenmanEtAl2006} to $\widetilde M_1$ and
$\widetilde M_2$ gives \eqref{eq:weakL1-limit} and
\eqref{eq:weakL1-tail}. For self-adjoint $A$, the assertion with $-i\eps$ follows by
applying the preceding result to $-A$.
\end{proof}

\begin{proof}[Proof of Lemma~\ref{lem:density}]
Let
\[
I_0:=(-1/2,1/2),
\]
and, for $k\ge1$, let
\[
I_{k,+}:=(1-2^{-k},1-2^{-k-1}),
\qquad
I_{k,-}:=(-1+2^{-k-1},-1+2^{-k}).
\]
Thus the intervals $I_0,I_{k,\pm}$ are pairwise disjoint and cover
$(-1,1)$ except for their endpoints. Writing $\ell_\nu:=|I_\nu|$, we have
\[
\ell_0=1,
\qquad
\ell_{k,\pm}=2^{-k-1},
\quad k\ge1.
\]

For an interval $I$, write $\frac32 I$ for the concentric interval with
length $\frac32|I|$. The closures of all $\frac32 I_\nu$ are contained in
$(-1,1)$, and at most three of these dilated intervals contain any given
point. Choose a fixed nonnegative
$\varphi\in C_c^\infty(-3/4,3/4)$ with $0\le\varphi\le1$ and
$\varphi=1$ on a neighborhood of $[-1/2,1/2]$. If $c_\nu$ is the center of
$I_\nu$, set
\[
b_\nu(t):=\varphi\!\left(\frac{t-c_\nu}{\ell_\nu}\right).
\]
Then $b_\nu=1$ on a neighborhood of $\overline{I_\nu}$,
\[
\supp b_\nu\subset\frac32 I_\nu,
\qquad
\|b_\nu'\|_\infty\le \|\varphi'\|_\infty\ell_\nu^{-1}.
\]
Since the $\overline{I_\nu}$ cover $(-1,1)$ and at most three
$b_\mu$ are nonzero at any point,
\[
1\le S(t):=\sum_\mu b_\mu(t)^2\le3,
\qquad t\in(-1,1).
\]
Set
\[
\chi_\nu:=\frac{b_\nu^2}{S}.
\]
Then $\chi_\nu\in C_c^\infty(-1,1)$, $\chi_\nu\ge0$, and
$\sum_\nu\chi_\nu=1$ on $(-1,1)$. Hence
\begin{equation}
c\ell_\nu\le\int\chi_\nu\le C\ell_\nu,
\qquad
\int_{\{\chi_\nu>0\}}\frac{|\chi_\nu'|^2}{\chi_\nu}
=4\int| (\sqrt{\chi_\nu})'|^2
\le C\ell_\nu^{-1}.
\label{eq:whitney-bounds}
\end{equation}

Let $Z_a:=\int_{-1}^1a$, and define
\[
A_\nu:=\int a\chi_\nu,\qquad
\pi_\nu:=\frac{A_\nu}{Z_a},\qquad
p_\nu:=\frac{a\chi_\nu}{A_\nu}.
\]
We have $q_a=\sum_\nu\pi_\nu p_\nu$ pointwise and
$\sum_\nu\pi_\nu=1$.
Since $0<c\le a\le C$, \eqref{eq:whitney-bounds} yields
$\pi_\nu\asymp\ell_\nu$. On $\{\chi_\nu>0\}$,
\begin{align*}
\frac{|(a\chi_\nu)'|^2}{a\chi_\nu}
&\le 2\frac{|a'|^2}{a}\chi_\nu
   +2a\frac{|\chi_\nu'|^2}{\chi_\nu},\\
I(p_\nu)&=\frac1{A_\nu}
\int_{\{\chi_\nu>0\}}\frac{|(a\chi_\nu)'|^2}{a\chi_\nu}
\le C_a\ell_\nu^{-2}.
\end{align*}
Consequently,
\[
\sum_\nu\pi_\nu I(p_\nu)^{1/3}
\le C_a\sum_{k\ge0}2^{-k/3}<\infty.
\]

Finally, let $b\in C_c^\infty(\bbR)$ be nonzero and real-valued, and set
$q=b^2/\|b\|_2^2$. Then
\[
I(q)=\frac{4}{\|b\|_2^2}\int_{\{b\ne0\}}|b'(t)|^2\,dt
\le\frac{4\|b'\|_2^2}{\|b\|_2^2}<\infty.
\]
\end{proof}

\begin{proof}[Proof of Lemma~\ref{lem:multi-site}]
Let $p$ be a probability density on $\bbR$ with $\sqrt p\in H^1(\bbR)$.
Then, in the distributional sense,
\begin{equation}
p'=2\sqrt p\,(\sqrt p)'.
\label{eq:weak-product-density}
\end{equation}
Cauchy--Schwarz then yields
\[
\|p'\|_{L^1}\le 2\|\sqrt p\|_{L^2}\|(\sqrt p)'\|_{L^2}<\infty.
\]
Since $p\in W^{1,1}(\bbR)$, it has a locally absolutely continuous
representative, still denoted by $p$, satisfying
\[
p(y)-p(x)=\int_x^y p'(t)\,dt,
\qquad x<y.
\]
By \eqref{eq:weak-product-density}, $p'=0$ almost everywhere on
$\{p=0\}$. Moreover, since $p,p'\in
L^1(\bbR)$, this representative has limits at $\pm\infty$, and both limits
are zero because $p\in L^1(\bbR)$. Hence $\int_{\bbR}p'=0$. Set
\begin{equation}
L_p(t):=\begin{cases}p'(t)/p(t),&p(t)>0,\\0,&p(t)=0.\end{cases}
\qquad
p'=pL_p,\quad
\int_{\bbR}L_p p=0,\quad
\int_{\bbR}L_p^2p=I(p).
\label{eq:score-identities}
\end{equation}

We first prove the estimate when each density $p_x$ has finite Fisher
information $I_x:=I(p_x)$. Let $V=(V_x)_{x\in X}$ denote the random
vector and write $v=(v_x)_{x\in X}\in\bbR^X$. Since $X$ is finite and the random variables $V_x$, $x\in X$, are independent, $V$ has
joint density
\[
P(v)=\prod_{x\in X}p_x(v_x).
\]
Fix a vector $w=(w_x)_{x\in X}$ with every $w_x>0$, and define the weak
directional derivative $D_wP:=\sum_xw_x\partial_{v_x}P$. By
\eqref{eq:score-identities} and independence,
\[
D_wP(v)=P(v)\sum_xw_xL_{p_x}(v_x)
\]
for Lebesgue-almost every $v\in\bbR^X$. By \eqref{eq:score-identities} and independence, Cauchy--Schwarz gives
\begin{equation}
\|D_wP\|_{L^1}
=\mathbb E\left|\sum_xw_xL_{p_x}(V_x)\right|
\le\left(\mathbb E\left|\sum_xw_xL_{p_x}(V_x)\right|^2\right)^{1/2}
=\left(\sum_xI_xw_x^2\right)^{1/2}.
\label{eq:directional-derivative}
\end{equation}

We next disintegrate the product measure along lines parallel to $w$. Write
$v=\xi+tw$ with $\xi\in w^\perp$ and $t\in\bbR$. Let $dv$ denote
Lebesgue measure on $\bbR^X$ and $d\xi$ the induced Lebesgue measure on
$w^\perp$. The corresponding change of variables has Jacobian $\|w\|_2$, so
\begin{equation}
dv=\|w\|_2\,d\xi\,dt.
\label{eq:line-jacobian}
\end{equation}
For almost every $\xi\in w^\perp$, the function
\begin{equation}
q_\xi(t):=\|w\|_2P(\xi+tw)
\label{eq:fiber-density}
\end{equation}
belongs to $W^{1,1}(\bbR)$ and satisfies
\[
q_\xi'(t)=\|w\|_2D_wP(\xi+tw)
\quad\text{for almost every }t\in\bbR.
\]
Its absolutely continuous representative is nonnegative and tends to zero
at both $\pm\infty$. Consequently
\[
q_\xi(t)
=\frac12\left(\int_{-\infty}^t q_\xi'(s)\,ds
-\int_t^\infty q_\xi'(s)\,ds\right),
\]
so
\[
\sup_tq_\xi(t)\le\frac12\int_{\bbR}|q_\xi'(t)|\,dt.
\]
By Fubini's theorem and \eqref{eq:line-jacobian},
\begin{align}
\int_{w^\perp}\sup_tq_\xi(t)\,d\xi
&\le\frac12\int_{w^\perp}\int_{\bbR}|q_\xi'(t)|\,dt\,d\xi\notag\\
&=\frac12\int_{w^\perp}\int_{\bbR}\|w\|_2|D_wP(\xi+tw)|\,dt\,d\xi\notag\\
&=\frac12\|D_wP\|_{L^1}.
\label{eq:line-density}
\end{align}

We now apply Lemma~\ref{lem:weakL1} on each line. Let
$U=\diag(w_x)$ and set $a_x=\|F\delta_x\|^2$. Since $U$ is positive and
invertible,
\begin{align*}
Q+\lambda\diag(\xi+tw)
&=Q+\lambda\diag\xi+\lambda tU\\
&=U^{1/2}(B_\xi+\lambda t)U^{1/2},
\end{align*}
where
\[
B_\xi:=U^{-1/2}(Q+\lambda\diag\xi)U^{-1/2}.
\]
Then
\[
\im B_\xi=U^{-1/2}(\im Q)U^{-1/2}<0.
\]
Thus $A_\xi:=-B_\xi$ is maximally dissipative. For $s\in\bbR$,
\begin{equation}
(B_\xi+s)^{-1}=-(A_\xi-s)^{-1}
=-\lim_{\eps\downarrow0}(A_\xi-s+i\eps)^{-1}.
\label{eq:line-boundary-identity}
\end{equation}
Moreover,
\begin{equation}
F(Q+\lambda\diag(\xi+tw))^{-1}F^*
=
FU^{-1/2}(B_\xi+\lambda t)^{-1}U^{-1/2}F^*.
\label{eq:line-resolvent-factor}
\end{equation}
\begin{equation}
\|FU^{-1/2}\|_{\HS}^2
=\sum_x\frac{a_x}{w_x}.
\label{eq:weighted-hs}
\end{equation}
By \eqref{eq:line-boundary-identity}, Lemma~\ref{lem:weakL1}, and
\eqref{eq:weighted-hs}, with $A_\xi=-B_\xi$,
$M_1=FU^{-1/2}$, and $M_2=U^{-1/2}F^*$,
\[
\Leb\left\{s:
\|FU^{-1/2}(B_\xi+s)^{-1}U^{-1/2}F^*\|_{\HS}>u\right\}
\le\frac{C}{u}\sum_x\frac{a_x}{w_x}.
\]
Using \eqref{eq:line-resolvent-factor} and the substitution
$s=\lambda t$ gives
\begin{equation}
\Leb\left\{t:
\|F(Q+\lambda\diag(\xi+tw))^{-1}F^*\|_{\HS}>u\right\}
\le\frac{C}{\lambda u}\sum_x\frac{a_x}{w_x}.
\label{eq:line-exception}
\end{equation}

Let $E_\xi(u):=\{t\in\bbR:\|F(Q+\lambda\diag(\xi+tw))^{-1}F^*\|_{\HS}>u\}$. Using \eqref{eq:fiber-density} and \eqref{eq:line-density},
\begin{align*}
&\mathbb P\left\{\|F(Q+\lambda\diag V)^{-1}F^*\|_{\HS}>u\right\}\\*
&\quad=\int_{w^\perp}\int_{E_\xi(u)}q_\xi(t)\,dt\,d\xi\\
&\quad\le
\frac{C}{\lambda u}\left(\sum_x\frac{a_x}{w_x}\right)
\int_{w^\perp}\sup_tq_\xi(t)\,d\xi\\
&\quad\le
\frac{C}{\lambda u}\|D_wP\|_{L^1}
\sum_x\frac{a_x}{w_x}.
\end{align*}
Combining this with \eqref{eq:directional-derivative} gives
\begin{equation}
\mathbb P\left\{\|F(Q+\lambda\diag V)^{-1}F^*\|_{\HS}>u\right\}
\le
\frac{C}{\lambda u}
\left(\sum_xI_xw_x^2\right)^{1/2}
\sum_x\frac{a_x}{w_x}.
\label{eq:tail-multi-site}
\end{equation}

It remains to optimize the right-hand side over the direction $w$. Suppose
first that $a_x>0$ and $I_x>0$ for every $x$. H\"older's inequality gives
\[
\sum_xa_x^{2/3}I_x^{1/3}
=\sum_x(I_xw_x^2)^{1/3}\left(\frac{a_x}{w_x}\right)^{2/3}
\le
\left(\sum_xI_xw_x^2\right)^{1/3}
\left(\sum_x\frac{a_x}{w_x}\right)^{2/3}.
\]
Thus,
\[
\left(\sum_xa_x^{2/3}I_x^{1/3}\right)^{3/2}
\le
\left(\sum_xI_xw_x^2\right)^{1/2}
\sum_x\frac{a_x}{w_x}.
\]
Equality holds when $w_x^3$ is proportional to $a_x/I_x$. Consequently,
\begin{equation}
\inf_{w_x>0}
\left(\sum_xI_xw_x^2\right)^{1/2}
\sum_x\frac{a_x}{w_x}
=
\left(\sum_xa_x^{2/3}I_x^{1/3}\right)^{3/2}.
\label{eq:weight-opt}
\end{equation}
By approximation, \eqref{eq:weight-opt} remains valid when some $a_x$ or
$I_x$ is zero.

Let
\[
X_F:=\|F(Q+\lambda\diag V)^{-1}F^*\|_{\HS}.
\]
By \eqref{eq:tail-multi-site} and \eqref{eq:weight-opt}, we have
$\mathbb P(X_F>u)\le\min(1,b/u)$ with
\[
b=C\lambda^{-1}
\left(\sum_xa_x^{2/3}I_x^{1/3}\right)^{3/2}.
\]
For $0<s<1$, the layer-cake formula gives
\begin{align*}
\mathbb EX_F^s
&=s\int_0^\infty u^{s-1}\mathbb P(X_F>u)\,du\\
&\le s\int_0^bu^{s-1}\,du
+sb\int_b^\infty u^{s-2}\,du
=\frac{b^s}{1-s}.
\end{align*}
Therefore
\begin{equation}
\mathbb EX_F^s
\le
C_s\lambda^{-s}
\left(\sum_xa_x^{2/3}I_x^{1/3}\right)^{3s/2}.
\label{eq:finite-fisher-moment}
\end{equation}

Finally we return to mixed Fisher information. For each $x$, write the chosen
decomposition as $p_x=\sum_\nu\pi_{x,\nu}p_{x,\nu}$, and introduce an
independent label $\nu_x$ with law $\pi_{x,\nu}$. Conditional on all labels,
the random variables $V_x$ remain independent and have densities $p_{x,\nu_x}$,
so \eqref{eq:finite-fisher-moment} applies with
$I_x=I(p_{x,\nu_x})$. Set $q=3s/2<1$. By Jensen's inequality,
\begin{align}
&\mathbb E_{\mathrm{labels}}
\left(\sum_xa_x^{2/3}I(p_{x,\nu_x})^{1/3}\right)^q\notag\\*
&\qquad\le
\left(
\sum_xa_x^{2/3}
\mathbb E_{\mathrm{labels}}I(p_{x,\nu_x})^{1/3}
\right)^q\notag\\
&\qquad\le
\left(M_*\sum_xa_x^{2/3}\right)^q
=M_*^{3s/2}\Phi(F)^s,
\label{eq:label-average}
\end{align}
Taking expectation over the labels in \eqref{eq:finite-fisher-moment}
and using \eqref{eq:label-average} yields \eqref{eq:multi-site}.
\end{proof}

\section{Proofs of the factorization estimates}
\label{app:factorization-proofs}

\begin{proof}[Proof of Lemma~\ref{lem:gamma-algebra}]
For each estimate it suffices to consider the case in which its right-hand
side is finite. Let $B=F^*G$ be any admissible factorization. By the Schatten--H\"older
inequality and \eqref{eq:phi-hs},
\[
\|B\|_{\TC}
=\|F^*G\|_{\TC}
\le\|F\|_{\HS}\|G\|_{\HS}
\le\sqrt{\Phi(F)\Phi(G)}.
\]
Taking the infimum over factorizations gives \eqref{eq:gamma-trace}.

To prove \eqref{eq:gamma-quasi}, first consider two fixed factorizations
$B_j=F_j^*G_j$. If $B_j=0$, use the zero factorization. Otherwise
$\Phi(F_j),\Phi(G_j)>0$. For $c_j>0$ set
\[
\widetilde F_j=c_jF_j,
\qquad
\widetilde G_j=c_j^{-1}G_j.
\]
Then $B_j=\widetilde F_j^*\widetilde G_j$. Since
$\Phi(cF)=c^2\Phi(F)$, the choice
\[
c_j=\left(\frac{\Phi(G_j)}{\Phi(F_j)}\right)^{1/4}
\]
gives
\[
\Phi(\widetilde F_j)=\Phi(\widetilde G_j)
=\sqrt{\Phi(F_j)\Phi(G_j)}.
\]
Let $\mathcal K_j$ be the auxiliary space of $\widetilde F_j,\widetilde G_j$,
set $\mathcal K=\mathcal K_1\oplus\mathcal K_2$, and define
\[
F u=(\widetilde F_1u,\widetilde F_2u),
\qquad
G v=(\widetilde G_1v,\widetilde G_2v).
\]
Then
\[
F^*G=\widetilde F_1^*\widetilde G_1+\widetilde F_2^*\widetilde G_2=B_1+B_2.
\]
Since $p_0=2/3<1$,
\[
\Phi(F)^{p_0}
=\sum_x\bigl(\|\widetilde F_1\delta_x\|^2+\|\widetilde F_2\delta_x\|^2\bigr)^{p_0}
\le\Phi(\widetilde F_1)^{p_0}+\Phi(\widetilde F_2)^{p_0}.
\]
The same bound holds for $G$. Writing
$a_j:=\sqrt{\Phi(F_j)\Phi(G_j)}
=\Phi(\widetilde F_j)=\Phi(\widetilde G_j)$, we get
\[
\gamma(B_1+B_2)^{p_0}
\le\bigl(\Phi(F)\Phi(G)\bigr)^{p_0/2}
\le a_1^{p_0}+a_2^{p_0}.
\]
Taking factorizations with $a_j$ arbitrarily close to $\gamma(B_j)$
proves \eqref{eq:gamma-quasi}.

The countable case follows by the same direct-sum construction, choosing
balanced factorizations with
$\sum_j a_j^{p_0}\le\sum_j\gamma(B_j)^{p_0}+\varepsilon$.
The summability assumption ensures that the series converges in $\mathcal J_1$
and that $\Phi(F),\Phi(G)<\infty$. The same estimate then applies,
and letting $\varepsilon\downarrow0$ proves \eqref{eq:gamma-countable}.

We next prove \eqref{eq:gamma-product}. Suppose $B=F^*G$ and set
$b_k:=\sup_n|A(n,n-k)|$. For every $y$,
\[
A\delta_y=\sum_n A(n,y)\delta_n
=\sum_k A(y+k,y)\delta_{y+k},
\]
so
\[
\|FA\delta_y\|
\le\sum_k b_k\|F\delta_{y+k}\|.
\]
By Minkowski's inequality,
\[
\left(\sum_y\|FA\delta_y\|^{4/3}\right)^{3/4}
\le
\sum_kb_k
\left(\sum_y\|F\delta_{y+k}\|^{4/3}\right)^{3/4}.
\]
Hence
\begin{equation}
\Phi(FA)\le\|A\|_{\mathcal C}^2\Phi(F).
\label{eq:phi-product}
\end{equation}
Similarly,
$\Phi(GC)\le\|C\|_{\mathcal C}^2\Phi(G)$. Finally
\[
ABC=(FA^*)^*(GC),
\]
so \eqref{eq:gamma-product} follows after taking the infimum over
factorizations of $B$. If $P_I,P_J$ are coordinate projections, then
\[
\gamma(P_JBP_I)
\le \|P_J\|_{\mathcal C}\gamma(B)\|P_I\|_{\mathcal C}
=\gamma(B).
\]

Finally, for \eqref{eq:gamma-finite}, write a singular-value decomposition
\[
B=\sum_{q=1}^N s_q\,|u_q\rangle\langle v_q|,
\]
where $N=\operatorname{rank}B$, $s_q>0$, $\{v_q\}\subset\ell^2(I)$ and
$\{u_q\}\subset\ell^2(J)$ are orthonormal families. Define
$L_I:\ell^2(I)\to\ell^2(\{1,\ldots,N\})$ and
$L_J:\ell^2(J)\to\ell^2(\{1,\ldots,N\})$ by
\[
L_I f=(s_q^{1/2}\langle v_q,f\rangle)_q,
\qquad
L_J g=(s_q^{1/2}\langle u_q,g\rangle)_q.
\]
Then
$L_J^*L_If=\sum_{q=1}^N s_q\langle v_q,f\rangle u_q=Bf$.
Moreover,
\[
\|L_I\|_{\HS}^2=\|L_J\|_{\HS}^2
=\sum_qs_q=\|B\|_{\TC}.
\]
If $c_x=\|L_I\delta_x\|^2$ for $x\in I$, H\"older's inequality gives
\[
\Phi(L_I)
=\left(\sum_{x\in I}c_x^{2/3}\right)^{3/2}
\le |I|^{1/2}\sum_{x\in I}c_x
=|I|^{1/2}\|B\|_{\TC}.
\]
Similarly,
$\Phi(L_J)\le|J|^{1/2}\|B\|_{\TC}$. Hence, by
Definition~\ref{def:gamma},
\[
\gamma(B)
\le\sqrt{\Phi(L_I)\Phi(L_J)}
\le(|I||J|)^{1/4}\|B\|_{\TC},
\]
which proves \eqref{eq:gamma-finite}.
\end{proof}

\begin{proof}[Proof of Lemma~\ref{lem:crosscut-product}]
Fix integers $x\le y$ and set
\[
m=\left\lfloor\frac{x+y}{2}\right\rfloor.
\]
Then
\[
1+m-x\ge\frac12(1+y-x),
\qquad
1+y-m\ge\frac12(1+y-x).
\]
Using $I=P_m^-+P_m^+$, we write
\begin{equation}
P_x^-ABP_y^+
=
P_x^-AP_m^-BP_y^+
+
P_x^-AP_m^+BP_y^+.
\label{eq:crosscut-split}
\end{equation}
Since $m\le y$ and $1+y-m\ge\frac12(1+y-x)$,
\eqref{eq:gamma-compression}, \eqref{eq:gamma-product}, and
\eqref{eq:crosscut} give
\begin{align*}
&(1+y-x)^\alpha\gamma(P_x^-AP_m^-BP_y^+)\\*
&\quad\le (1+y-x)^\alpha\|A\|_{\mathcal C}
\gamma(P_m^-BP_y^+)\\
&\quad\le 2^\alpha\|A\|_{\mathcal C}
(1+y-m)^\alpha\gamma(P_m^-BP_y^+)\\
&\quad\le 2^\alpha\|A\|_{\mathcal C}g_\alpha(B).
\end{align*}
Similarly,
\[
(1+y-x)^\alpha\gamma(P_x^-AP_m^+BP_y^+)
\le 2^\alpha g_\alpha(A)\|B\|_{\mathcal C}.
\]
Applying \eqref{eq:gamma-quasi} to \eqref{eq:crosscut-split}, we obtain
\[
(1+y-x)^{\alpha p_0}\gamma(P_x^-ABP_y^+)^{p_0}
\le 2^{\alpha p_0}\left(
\|A\|_{\mathcal C}^{p_0}g_\alpha(B)^{p_0}
+g_\alpha(A)^{p_0}\|B\|_{\mathcal C}^{p_0}
\right).
\]
The same argument applies to $P_y^+ABP_x^-$.
Taking the supremum over $x\le y$ proves \eqref{eq:crosscut-product}.
\end{proof}

\section{Proofs of the resolvent and coupling estimates}
\label{app:resolvent-proofs}

\begin{proof}[Proof of Lemma~\ref{lem:pointwise-algebra}]
For \eqref{eq:C-J}, fix $k\in\bbZ$. Then
\[
\sup_n|A(n,n-k)|\le \langle k\rangle^{-r}\|A\|_{J_r}.
\]
Summing over $k$ and using $r>1$ gives
\[
\|A\|_{\mathcal C}
=\sum_k\sup_n|A(n,n-k)|
\le \|A\|_{J_r}\sum_k\langle k\rangle^{-r}
\le C_r\|A\|_{J_r}.
\]

For \eqref{eq:S-interp}, the case $A=0$ is immediate. If
$0<\|A\|_{J_r}<\|A\|$, then
\[
S(A)\le C_r\|A\|_{J_r}
\le C_r\|A\|^{\frac{2r-2}{2r-1}}\|A\|_{J_r}^{\frac1{2r-1}}.
\]
We may therefore assume $0<\|A\|\le\|A\|_{J_r}$.
For $R\ge1$, the Cauchy--Schwarz inequality gives
\begin{align*}
\sup_x\sum_y|A(x,y)|
&=\sup_x\left(\sum_{|x-y|\le R}|A(x,y)|+\sum_{|x-y|>R}|A(x,y)|\right)\\
&\le (2R+1)^{1/2}\sup_x\left(\sum_y|A(x,y)|^2\right)^{1/2}
+\|A\|_{J_r}\sum_{|k|>R}\langle k\rangle^{-r}\\
&=(2R+1)^{1/2}\sup_x\|A^*\delta_x\|_2
+\|A\|_{J_r}\sum_{|k|>R}\langle k\rangle^{-r}\\
&\le C_r\left(R^{1/2}\|A\|+R^{1-r}\|A\|_{J_r}\right).
\end{align*}
Applying the same estimate to $A^*$, we obtain
\begin{equation}
S(A)\le C_r\left(R^{1/2}\|A\|+R^{1-r}\|A\|_{J_r}\right).
\label{eq:S-before-balance}
\end{equation}
Taking
\[
R=\left(\frac{\|A\|_{J_r}}{\|A\|}\right)^{2/(2r-1)}\ge1
\]
in \eqref{eq:S-before-balance} proves \eqref{eq:S-interp}.

Finally,
\begin{align*}
\langle x-y\rangle^r|(AB)(x,y)|
&\le\sum_z\langle x-y\rangle^r|A(x,z)|\,|B(z,y)|\\
&\le C_r\sum_z\left(\langle x-z\rangle^r+\langle z-y\rangle^r\right)
|A(x,z)|\,|B(z,y)|\\
&\le C_r\left(\|A\|_{J_r}\sum_z|B(z,y)|
+\|B\|_{J_r}\sum_z|A(x,z)|\right)\\
&\le C_r\left(\|A\|_{J_r}S(B)+S(A)\|B\|_{J_r}\right).
\end{align*}
Taking the supremum over $x,y$ proves \eqref{eq:J-product}.
\end{proof}

\begin{lemma}\label{lem:contraction-powers}
Let $B\in\mathcal B(\ell^2(\bbZ))$ satisfy $\|B\|\le q<1$, $\|B\|_{J_r}<\infty$, and
$g_\alpha(B)<\infty$. Then, for every $q<q_1<1$, there exist constants
$C,M>0$ such that
\begin{equation}
\|B^k\|_{J_r}+\|B^k\|_{\mathcal C}\le Cq_1^k,
\qquad
g_\alpha(B^k)\le Ck^Mq_1^k,
\qquad k\ge1.
\label{eq:contraction-powers}
\end{equation}
\end{lemma}

\begin{proof}
If $q=0$, then $B=0$ and both estimates hold for every $k\ge1$.
Thus we may assume $0<q<1$.
Set
\[
p=1+\frac1{2r-1}\in(1,2),
\qquad
\delta=\log_2p\in(0,1).
\]
By \eqref{eq:J-product} and \eqref{eq:S-interp},
\[
\|A^2\|_{J_r}
\le 2C_r\|A\|_{J_r}S(A)
\le C_r\|A\|^{\frac{2r-2}{2r-1}}
       \|A\|_{J_r}^{1+\frac1{2r-1}}.
\]
Since $2-p=(2r-2)/(2r-1)$, this is
\begin{equation}
\|A^2\|_{J_r}
\le C_r\|A\|^{2-p}\|A\|_{J_r}^p.
\label{eq:square-recursion}
\end{equation}

Let $X_j=\|B^{2^j}\|_{J_r}$. $\|B^{2^j}\|\le\|B\|^{2^j}\le q^{2^j}$, so
\[
X_{j+1}\le C_rq^{(2-p)2^j}X_j^p.
\]
If $X_j=0$ for some $j$, then $B^k=0$ for all $k\ge2^j$,
and \eqref{eq:contraction-powers} follows by enlarging $C$.
Hence we may assume $X_j>0$ for every $j$ and define
\[
Y_j=\log X_j-2^j\log q.
\]
Then
\[
Y_{j+1}\le\log C_r+pY_j.
\]
Iterating this inequality,
\[
Y_j
\le p^jY_0+\log C_r\sum_{\ell=0}^{j-1}p^\ell
\le Cp^j.
\]
Hence
\begin{equation}
X_j\le q^{2^j}\exp(Cp^j)
=q^{2^j}\exp(C2^{j\delta}).
\label{eq:dyadic-power-bound}
\end{equation}

For a general $k$, write its binary expansion
$k=\sum_{\ell=1}^N2^{j_\ell}$ with distinct nonnegative integers $j_\ell$.
Then $N\le1+\log_2k$. If $J=\max_\ell j_\ell$, then $2^J\le k$ and
\[
\sum_{\ell=1}^N2^{j_\ell\delta}
\le\sum_{j=0}^J2^{j\delta}
\le C_\delta2^{J\delta}
\le C_\delta k^\delta.
\]
Set $D_\ell=B^{2^{j_\ell}}$, so $B^k=D_1\cdots D_N$. From
\eqref{eq:J-product} and \eqref{eq:C-J},
\[
\|AB\|_{J_r}\le C_*\|A\|_{J_r}\|B\|_{J_r}.
\]
Here $C_*\ge1$ depends only on $r$.
Since $0<\delta<1$ and $q<q_1<1$, iterating this product estimate and applying \eqref{eq:dyadic-power-bound} to
all $D_\ell$ gives
\begin{align*}
\|B^k\|_{J_r}
&\le C_*^{N-1}\prod_{\ell=1}^N\|D_\ell\|_{J_r}\\
&\le C_*^{N-1}
\prod_{\ell=1}^N
\left(q^{2^{j_\ell}}\exp(C2^{j_\ell\delta})\right)\\
&=C_*^{N-1}q^k
\exp\left(C\sum_{\ell=1}^N2^{j_\ell\delta}\right)\\
&\le q^k\exp(C'k^\delta+C'\log(1+k))\\
&\le Cq_1^k,\qquad k\ge1.
\end{align*}
By \eqref{eq:C-J}, $\|B^k\|_{\mathcal C}\le Cq_1^k$.

Set $H_k=g_\alpha(B^k)^{p_0}$. Since
$\|B^m\|_{\mathcal C}\le Cq_1^m$,
Lemma~\ref{lem:crosscut-product}, applied to
$B^{m+n}=B^mB^n$, gives
\[
H_{m+n}\le C\bigl(q_1^{p_0m}H_n+q_1^{p_0n}H_m\bigr).
\]
Set $Z_k=q_1^{-p_0k}H_k$. Then
\begin{equation}
Z_{m+n}\le C(Z_m+Z_n).
\label{eq:Z-recursion}
\end{equation}
For $L\ge0$, set $A_L=\max_{1\le n\le2^L}Z_n$. For
$2\le n\le2^L$, set $m=\lfloor n/2\rfloor$ and
$\ell=\lceil n/2\rceil$. Then $n=m+\ell$ and
$1\le m,\ell\le2^{L-1}$, so $Z_m,Z_\ell\le A_{L-1}$. Thus, by
\eqref{eq:Z-recursion},
\[
Z_n=Z_{m+\ell}\le C(Z_m+Z_\ell)\le2CA_{L-1}.
\]
Thus $A_L\le C_1A_{L-1}$ with $C_1=\max\{1,2C\}$, so
$A_L\le A_0C_1^L$. Choosing $L=\lceil\log_2k\rceil$, we have
$k\le2^L$ and
\[
Z_k\le A_L\le A_0C_1^L
\le A_0C_1\,k^{\log_2C_1}.
\]
Thus
\[
g_\alpha(B^k)=q_1^kZ_k^{1/p_0}
\le(A_0C_1)^{1/p_0}k^{(\log_2C_1)/p_0}q_1^k
\le Ck^Mq_1^k,\qquad k\ge1,
\]
where $M=\max\{1,(\log_2C_1)/p_0\}$.
\end{proof}

\begin{proof}[Proof of Proposition~\ref{prop:uniform-inverse}]
On $\ell^2(\bbZ)$, define
\[
Q:=P_\Lambda(H_\Lambda-z)P_\Lambda+P_{\Lambda^c}
=T_{r,\Lambda}+D,
\qquad D:=P_\Lambda(v-z)P_\Lambda+P_{\Lambda^c}.
\]
Then
\[
\|T_r\|_{J_r}=2^r,\qquad
\|D\|_{J_r}=\|D\|_{\mathcal C}=\|D\|
\le\max\{V_0+E_0,1\},\qquad g_\alpha(D)=0.
\]
Consequently,
\[
\|Q\|_{J_r}\le \|T_{r,\Lambda}\|_{J_r}+\|D\|_{J_r}
\le\|T_r\|_{J_r}+\|D\|_{J_r}\le C.
\]
By \eqref{eq:C-J}, $\|Q\|_{\mathcal C}\le C_r\|Q\|_{J_r}$ and
$\|Q\|\le\|Q\|_{\mathcal C}$. By \eqref{eq:gamma-compression},
\[
g_\alpha(Q)=g_\alpha(T_{r,\Lambda})
\le g_\alpha(T_r)\le C_r.
\]
Thus
\begin{equation}
\|Q\|+\|Q\|_{J_r}+\|Q\|_{\mathcal C}+g_\alpha(Q)
\le C_Q,\qquad C_Q=C_Q(r,V_0,E_0)<\infty.
\label{eq:Q-initial-bounds}
\end{equation}

Because $T_{r,\Lambda}$ is self-adjoint and $v$ is real-valued,
$H_\Lambda$ is self-adjoint. The spectral theorem therefore gives
\[
\|(H_\Lambda-z)^{-1}\|
=\dist(z,\sigma(H_\Lambda))^{-1}\le\delta^{-1}.
\]
\[
\|Q^{-1}\|\le\max\{\|(H_\Lambda-z)^{-1}\|,1\}
\le\max\{\delta^{-1},1\}.
\]
Hence $Q^*Q\ge cI$, where $c=\min\{1,\delta^2\}$. Choose
$a>C_Q^2$, and set
\[
B=I-a^{-1}Q^*Q.
\]
Since
\[
cI\le Q^*Q\le\|Q\|^2I<aI,
\]
we get
\[
0\le B\le\left(1-\frac{c}{a}\right)I=:qI,
\qquad q<1.
\]
By \eqref{eq:J-product} and \eqref{eq:C-J},
\[
\|Q^*Q\|_{J_r}
\le C_r\left(\|Q^*\|_{J_r}S(Q)+S(Q^*)\|Q\|_{J_r}\right)
\le2C_r\|Q\|_{J_r}\|Q\|_{\mathcal C}\le C.
\]
Also, \eqref{eq:C-J} gives $\|Q^*Q\|_{\mathcal C}\le C$.
By Lemma~\ref{lem:crosscut-product},
\[
g_\alpha(Q^*Q)^{p_0}
\le C_\alpha\left(\|Q^*\|_{\mathcal C}^{p_0}g_\alpha(Q)^{p_0}
+g_\alpha(Q^*)^{p_0}\|Q\|_{\mathcal C}^{p_0}\right)\le C.
\]
Hence
\[
\begin{aligned}
\|B\|_{J_r}+\|B\|_{\mathcal C}
&\le2+a^{-1}\left(\|Q^*Q\|_{J_r}+\|Q^*Q\|_{\mathcal C}\right)\le C,\\
g_\alpha(B)&=a^{-1}g_\alpha(Q^*Q)\le C.
\end{aligned}
\]
By the Neumann series,
\begin{equation}
Q^{-1}=(Q^*Q)^{-1}Q^*=a^{-1}(I-B)^{-1}Q^*
=a^{-1}\sum_{k\ge0}B^kQ^*.
\label{eq:inverse-series}
\end{equation}
Fix $q<q_1<1$. By Lemma~\ref{lem:contraction-powers},
\[
\|B^k\|_{J_r}+\|B^k\|_{\mathcal C}\le Cq_1^k,\qquad k\ge0.
\]
By \eqref{eq:J-product} and \eqref{eq:C-J},
\[
\|B^kQ^*\|_{J_r}
\le C_r\left(\|B^k\|_{J_r}S(Q^*)+S(B^k)\|Q^*\|_{J_r}\right)
\le Cq_1^k.
\]
Consequently,
\[
\begin{aligned}
\|Q^{-1}\|_{J_r}+\|Q^{-1}\|_{\mathcal C}
&\le a^{-1}\sum_{k\ge0}\left(\|B^kQ^*\|_{J_r}+\|B^kQ^*\|_{\mathcal C}\right)\\
&\le a^{-1}(1+C_r)\sum_{k\ge0}\|B^kQ^*\|_{J_r}
\le C\sum_{k\ge0}q_1^k\le C.
\end{aligned}
\]
By Lemmas~\ref{lem:crosscut-product} and \ref{lem:contraction-powers},
\[
\begin{aligned}
g_\alpha(B^kQ^*)^{p_0}
&\le C_\alpha\left(\|B^k\|_{\mathcal C}^{p_0}g_\alpha(Q^*)^{p_0}
+g_\alpha(B^k)^{p_0}\|Q^*\|_{\mathcal C}^{p_0}\right)\\
&\le C\left(q_1^{kp_0}+(1+k)^{Mp_0}q_1^{kp_0}\right)
\le C(1+k)^{Mp_0}q_1^{kp_0},\qquad k\ge0.
\end{aligned}
\]
For $x\le y$ and either $(P,P')=(P_x^-,P_y^+)$ or
$(P,P')=(P_y^+,P_x^-)$, \eqref{eq:gamma-countable} gives
\[
\begin{aligned}
\gamma(PQ^{-1}P')^{p_0}
&\le a^{-p_0}\sum_{k\ge0}\gamma(PB^kQ^*P')^{p_0}\\
&\le a^{-p_0}(1+y-x)^{-\alpha p_0}
\sum_{k\ge0}g_\alpha(B^kQ^*)^{p_0}.
\end{aligned}
\]
Thus
\[
g_\alpha(Q^{-1})^{p_0}
\le C\sum_{k\ge0}(1+k)^{Mp_0}q_1^{kp_0}\le C.
\]
Since
\[
(H_\Lambda-z)^{-1}=P_\Lambda Q^{-1}P_\Lambda\big|_{\ell^2(\Lambda)},
\]
\eqref{eq:gamma-compression} and \eqref{eq:C-J} give
\[
\begin{aligned}
&\|(H_\Lambda-z)^{-1}\|_{J_r}+\|(H_\Lambda-z)^{-1}\|_{\mathcal C}
+g_\alpha((H_\Lambda-z)^{-1})\\
&\qquad\le(1+C_r)\|Q^{-1}\|_{J_r}+g_\alpha(Q^{-1})\le C.
\end{aligned}
\]
\end{proof}

\begin{proof}[Proof of Corollary~\ref{cor:weighted-J}]
Define the unitary operator $W_\tau:\ell^2_\tau(\bbZ)\to\ell^2(\bbZ)$ by
\[
(W_\tau u)(n):=\langle n\rangle^\tau u(n).
\]
Since $\langle x\rangle/\langle y\rangle\le\langle x-y\rangle$,
\[
|(W_\tau AW_\tau^{-1})(x,y)|
=\frac{\langle x\rangle^\tau}{\langle y\rangle^\tau}|A(x,y)|
\le\|A\|_{J_r}\langle x-y\rangle^{-(r-\tau)}.
\]
Schur's test gives
\[
\|A\|_{\mathcal B(\ell^2_\tau)}
=\|W_\tau AW_\tau^{-1}\|_{\mathcal B(\ell^2)}
\le\|A\|_{J_r}\sum_{k\in\bbZ}\langle k\rangle^{-(r-\tau)}
\le C_{r,\tau}\|A\|_{J_r}.
\]
\end{proof}

\begin{lemma}\label{lem:exterior-column}
Let $I=[a,b]\cap\bbZ$ with $a,b\in\bbZ$, $a\le b$, and set
$N_I=|I|$ and $a_I(x)=\|P_IT_r\delta_x\|_2$ for $x\notin I$. Then
\begin{align}
a_I(x)&\le C_r\dist(x,I)^{1/2-r},\label{eq:aI-tail}\\
a_I(x)&\le\sqrt{N_I}\,\dist(x,I)^{-r},\label{eq:aI-volume}\\
\sum_{x\notin I}a_I(x)&\le C_r.\label{eq:aI-sum}
\end{align}
The last estimate uses $r>3/2$.
\end{lemma}

\begin{proof}
Write $d_x=\dist(x,I)\ge1$. Then
\[
a_I(x)^2
=\sum_{u\in I}|u-x|^{-2r}
\le\sum_{k\ge d_x}k^{-2r}
\le C_r d_x^{1-2r},
\]
which gives \eqref{eq:aI-tail}. Also,
\[
a_I(x)^2
=\sum_{u\in I}|u-x|^{-2r}
\le N_Id_x^{-2r},
\]
which proves \eqref{eq:aI-volume}.

For each $k\ge1$ there are at most two sites outside $I$ at distance $k$ from
$I$.  For $r>3/2$, \eqref{eq:aI-tail} gives
\[
\sum_{x\notin I}a_I(x)
\le2C_r\sum_{k\ge1}k^{1/2-r}\le C_r.
\]

\end{proof}

\begin{proof}[Proof of Proposition~\ref{prop:rectangular-self-energy}]
For $x\in B$, set $u_x=P_IT_r\delta_x\in\ell^2(I)$, so
$\|u_x\|=a_I(x)$. For $y\in B$, set
$v_y=P_JT_r\delta_y\in\ell^2(J)$, so $\|v_y\|=a_J(y)$.
For every $x,y\in B$,
\[
d\le\dist(x,I)+|x-y|+\dist(y,J).
\]
By \eqref{eq:aI-volume} and \eqref{eq:aI-sum},
\begin{align}
&\|T_{IB}RT_{BJ}\|_{\TC}\notag\\*
&=\left\|\sum_{x,y\in B}R(x,y)|u_x\rangle\langle v_y|\right\|_{\TC}\notag\\
&\le\sum_{x,y\in B}|R(x,y)|\,\|u_x\|\,\|v_y\|\notag\\
&=\sum_{x,y\in B}|R(x,y)|a_I(x)a_J(y)\notag\\
&\le\|R\|_{J_r}\sum_{x,y\in B}a_I(x)\langle x-y\rangle^{-r}a_J(y)\notag\\
&\le\|R\|_{J_r}\Biggl[
\sum_{\substack{x,y\in B\\\dist(x,I)\ge d/3}}
a_I(x)\langle x-y\rangle^{-r}a_J(y)
+\sum_{\substack{x,y\in B\\|x-y|\ge d/3}}
a_I(x)\langle x-y\rangle^{-r}a_J(y)\notag\\*
&\qquad+\sum_{\substack{x,y\in B\\\dist(y,J)\ge d/3}}
a_I(x)\langle x-y\rangle^{-r}a_J(y)\Biggr]\notag\\
&\le C_r\|R\|_{J_r}d^{-r}\left[
\sqrt{N_I}\sum_{y\in B}a_J(y)
+\left(\sum_{x\in B}a_I(x)\right)\left(\sum_{y\in B}a_J(y)\right)\right.\notag\\*
&\hspace{5em}\left.+\sqrt{N_J}\sum_{x\in B}a_I(x)\right]\notag\\
&\le C_r\|R\|_{J_r}\left(\sqrt{N_I}+1+\sqrt{N_J}\right)d^{-r}
\le C_r\|R\|_{J_r}\sqrt{N_IN_J}\,d^{-r}.
\label{eq:self-energy-rankone}
\end{align}
Assume without loss of generality that $I$ lies to the left
of $J$, and set
\[
a=\max I,\qquad b=\min J,\qquad d=b-a.
\]
For $i\in I$ and $j\in J$, set
\[
p=a-i\ge0,\qquad q=j-b\ge0.
\]
Then $|i-j|=d+p+q$. By \eqref{eq:laplace-power},
\[
T_{IJ}(i,j)
=\frac1{\Gamma(r)}\int_0^\infty
t^{r-1}e^{-dt}e^{-(a-i)t}e^{-(j-b)t}\,dt.
\]
Define
\[
u_I(t)(i)=e^{-(a-i)t},\qquad
u_J(t)(j)=e^{-(j-b)t}.
\]
Thus
\[
T_{IJ}
=\frac1{\Gamma(r)}\int_0^\infty t^{r-1}e^{-dt}
|u_I(t)\ra\la u_J(t)|\,dt.
\]
Moreover,
\[
\|u_I(t)\|^2=\sum_{i\in I}e^{-2(a-i)t}\le N_I,
\qquad
\|u_J(t)\|^2=\sum_{j\in J}e^{-2(j-b)t}\le N_J.
\]
Therefore
\[
\|T_{IJ}\|_{\TC}
\le \frac{\sqrt{N_IN_J}}{\Gamma(r)}
\int_0^\infty t^{r-1}e^{-dt}\,dt
=\sqrt{N_IN_J}\,d^{-r}.
\]
Combining this with \eqref{eq:rectangular-self-energy} proves
\eqref{eq:V-two-bounds}.

Set $A:=T_rRT_r$.
By Proposition~\ref{prop:uniform-inverse},
\[
\|R\|_{J_r}+\|R\|_{\mathcal C}+g_\alpha(R)\le C.
\]
Since $\|T_r\|_{\mathcal C}<\infty$ and $g_\alpha(T_r)<\infty$,
Lemma~\ref{lem:crosscut-product} first gives
\[
g_\alpha(T_rR)^{p_0}
\le C_\alpha\bigl(\|T_r\|_{\mathcal C}^{p_0}g_\alpha(R)^{p_0}
+g_\alpha(T_r)^{p_0}\|R\|_{\mathcal C}^{p_0}\bigr)\le C.
\]
By \eqref{eq:J-product} and \eqref{eq:C-J},
$\|T_rR\|_{\mathcal C}\le C$.
Applying Lemma~\ref{lem:crosscut-product} again gives $g_\alpha(A)\le C$.
By \eqref{eq:gamma-quasi}, \eqref{eq:gamma-compression}, and
\eqref{eq:crosscut},
\begin{align*}
\gamma(V_{IJ}(z))^{p_0}
&\le\gamma(P_IT_rP_J)^{p_0}+\gamma(P_IAP_J)^{p_0}\\
&\le\gamma(P_a^-T_rP_{b-1}^+)^{p_0}
+\gamma(P_a^-AP_{b-1}^+)^{p_0}\\
&\le\bigl(g_\alpha(T_r)^{p_0}+g_\alpha(A)^{p_0}\bigr)d^{-\alpha p_0}\\
&\le Cd^{-\alpha p_0}.
\end{align*}
Taking the $p_0$-th root proves \eqref{eq:V-gamma-bound}.
\end{proof}

\section{Measurability}
\label{app:measurability}
\begin{lemma}\label{lem:measurable-balanced}
Fix $m,n\ge1$.

(i) There exist Borel maps
\[
F_1:\bbC^{m\times n}\to\bbC^{(m+n)\times m},
\qquad
G_1:\bbC^{m\times n}\to\bbC^{(m+n)\times n}
\]
such that
\[
F_1(A)^*G_1(A)=A,\qquad
\Phi(F_1(A))=\Phi(G_1(A))\le2\gamma(A).
\]

(ii) There exist Borel maps
\[
F_2:\bbC^{m\times n}\to\bbC^{(m+n)\times m},
\qquad
G_2:\bbC^{m\times n}\to\bbC^{(m+n)\times n}
\]
such that
\[
F_2(A)^*G_2(A)=A,\qquad
\Phi(F_2(A))=\Phi(G_2(A))
\le(mn)^{1/4}\|A\|_{\TC}.
\]

In both cases, $F_\ell(0)=G_\ell(0)=0$.

\end{lemma}
\begin{proof} For a positive semidefinite matrix $X=(X_{xy})_{x,y=1}^m$, set
\[
\phi_m(X):=\left(\sum_{x=1}^mX_{xx}^{2/3}\right)^{3/2},
\]
and define $\phi_n(Y)$ analogously. Since $(F^*F)_{xx}=\|Fe_x\|^2$, we have
\[
\phi_m(F^*F)=\Phi(F).
\]
For $T\ge0$, define
\[
\mathcal M(A,T):=
\left\{
\begin{pmatrix}X&A\\A^*&Y\end{pmatrix}\ge0:
\phi_m(X)=\phi_n(Y)\le T
\right\}.
\]
Set
\[
T_1(A):=2\gamma(A),\qquad
T_2(A):=(mn)^{1/4}\|A\|_{\TC}.
\]

We first verify that $\mathcal M(A,T_\ell(A))$ is nonempty. For any factorization $A=F^*G$,
\[
|A_{xy}|\le\|Fe_x\|\|Ge_y\|
\le\sqrt{\Phi(F)\Phi(G)}.
\]
Thus $\gamma(A)>0$ whenever $A\ne0$. By Definition~\ref{def:gamma}, such an $A$ admits a factorization satisfying
\[
\sqrt{\Phi(F)\Phi(G)}<2\gamma(A).
\]
By rescaling $F$ and $G$, we may assume that $\Phi(F)=\Phi(G)<T_1(A)$. Then
\[
\begin{pmatrix}F^*F&A\\A^*&G^*G\end{pmatrix}\in\mathcal M(A,T_1(A)).
\]
For the second bound, the construction in the proof of \eqref{eq:gamma-finite}, followed by rescaling, gives $A=F^*G$ with $\Phi(F)=\Phi(G)\le T_2(A)$. Hence
\[
\begin{pmatrix}F^*F&A\\A^*&G^*G\end{pmatrix}\in\mathcal M(A,T_2(A)).
\]
At $A=0$, take the zero matrix.

By \eqref{eq:gamma-quasi} and \eqref{eq:gamma-finite},
\[
\bigl|\gamma(A)^{p_0}-\gamma(B)^{p_0}\bigr|
\le\gamma(A-B)^{p_0}
\le(mn)^{p_0/4}\|A-B\|_{\TC}^{p_0}.
\]
Consequently, $\gamma$, $T_1$, and $T_2$ are continuous.

Fix $\ell\in\{1,2\}$ and $A\in\bbC^{m\times n}$. The set $\mathcal M(A,T_\ell(A))$ is closed.
Since $X_{xx}\le\phi_m(X)$ and $Y_{yy}\le\phi_n(Y)$, we have
\[
X_{xx},Y_{yy}\le T_\ell(A).
\]
Moreover,
\[
|X_{xx'}|^2\le X_{xx}X_{x'x'}\le T_\ell(A)^2,
\qquad |Y_{yy'}|^2\le Y_{yy}Y_{y'y'}\le T_\ell(A)^2.
\]
Thus $\mathcal M(A,T_\ell(A))$ is compact.
Moreover, if $A_j\to A$, $M_j\to M$, and $M_j\in\mathcal M(A_j,T_\ell(A_j))$, then $M\in\mathcal M(A,T_\ell(A))$, since positive semidefiniteness is preserved under limits and $\phi_m,\phi_n,T_\ell$ are continuous.
The Arsenin–Kunugui theorem \cite[Theorem~18.18]{Kechris1995CDST} yields a Borel selection
\[
A\longmapsto M_\ell(A)\in\mathcal M(A,T_\ell(A)).
\]

Let $J_m u=(u,0)$ and $J_n v=(0,v)$ be the canonical embeddings into $\bbC^{m+n}$. Define
\[
F_\ell(A):=M_\ell(A)^{1/2}J_m,\qquad
G_\ell(A):=M_\ell(A)^{1/2}J_n.
\]
These maps are Borel. Writing
\[
M_\ell(A)=
\begin{pmatrix}X&A\\A^*&Y\end{pmatrix},
\]
we obtain
\[
F_\ell(A)^*G_\ell(A)=A,\qquad
\Phi(F_\ell(A))=\phi_m(X)=\phi_n(Y)
=\Phi(G_\ell(A))\le T_\ell(A).
\]
Finally, $T_\ell(0)=0$ forces $M_\ell(0)=0$, so $F_\ell(0)=G_\ell(0)=0$.
\end{proof}

\section*{Competing interests}
The authors declare that they have no conflict of interest.

\section*{Financial support}
The work of Y. Shi and H. Jiang was supported in part by the National Natural Science Foundation of China (Grant No. 12522110).

\section*{Data availability}
The manuscript has no associated data.

\section*{AI statement}
The authors declare that the main ideas of this paper are their own. ChatGPT 6.1 sol (OpenAI) was used to assist with literature searches, language editing, checking calculations and mathematical arguments, and preparing Figure~\ref{fig:barriers}. The authors reviewed and verified all AI-assisted material included in the paper and take full responsibility for the accuracy, originality, and integrity of the entire paper.

\bibliographystyle{localalpha}
\bibliography{references}

\end{document}